\documentclass[submitting]{nst}

\usepackage{subfigure,dcolumn}
\usepackage[T2A,T1]{fontenc}
\usepackage[russian,english]{babel}
\usepackage{multirow}
\usepackage{booktabs}
\usepackage{listings}
\usepackage{color}

\begin{document}
	
	\title{A method for sharing dynamic geometry information in studies on liquid-based detectors}
	\thanks{Supported by the National Natural Science Foundation of China (No. 11675275, 11975021, 
		U1932101) and the Strategic Priority Research Program of Chinese Academy of Sciences (XDA10010900).}

	\author{Shu Zhang}
	\affiliation{School of Physics, Sun Yat-Sen University, Guangzhou 510275, China}
	\author{Jing-Shu Li}
	\affiliation{School of Physics, Sun Yat-Sen University, Guangzhou 510275, China}
	\author{Yang-Jie Su}
	\affiliation{School of Physical Sciences, University of Chinese Academy of Sciences, Beijing 100049, China}
	\author{Yu-Mei Zhang}
	\email[Yu-Mei Zhang ]{zhangym26@mail.sysu.edu.cn}
	\affiliation{Sino-French Institute of Nuclear Engineering and Technology, Sun Yat-Sen University, Zhuhai 519082, China}
	\author{Zi-Yuan Li}
	\affiliation{School of Physics, Sun Yat-Sen University, Guangzhou 510275, China}
	\author{Zheng-Yun You}
	\email[Zheng-Yun You ]{youzhy5@mail.sysu.edu.cn}
	\affiliation{School of Physics, Sun Yat-Sen University, Guangzhou 510275, China}
	
	\begin{abstract}
		Liquid-based detectors are widely used in particle and nuclear physics experiments. Because fixed method is used to construct the geometry in detector simulations such as Geant4, it is usually difficult to describe the non-uniformity of the liquid in a detector. We propose a method based on geometry description markup language and a tessellated detector description to share the detector geometry information between computational fluid dynamics simulation software and detector simulation software. This method makes it possible to study the impact of a liquid flow and non-uniformity on the key performance of a liquid-based detector, such as the event vertex reconstruction resolution. This will also be helpful in the detector design and performance optimization.
	\end{abstract}
	
	\keywords{Liquid-based detector, Geometry, Simulation, Geant4, Computational fluid dynamics}
	
	\maketitle
	
	\section{Introduction}
	
	In particle and nuclear physics experiments, large-scale liquid-based detectors have been widely used, particularly in neutrino detectors and in experiments searching for dark matter or neutrino-less double-beta decay. For example, the Super-Kamiokande detector used 50,000 tons of ultra-pure water to detect neutrinos\cite{abe2013evidence}. In addition, Sudbury Neutrino Observatory used 1000 tons of heavy water to detect solar neutrinos\cite{adam2015juno}. In the Jiangmen Underground Neutrino Observatory (JUNO), 20,000 tons of linear alkylbenzene liquid scintillator will be used to detect the reactor neutrinos\cite{adam2015juno}. For the two future flagship neutrino experiments, Hyper-Kamiokande\cite{Abe2018uyc} and the Deep Underground Neutrino Experiment (DUNE)\cite{acciarri2016long} will use 1 million tons of ultrapure water and 40,000 tons of liquid argon, respectively. In addition to neutrino experiments, liquid-based detectors have been used in dark matter studies, such as the Particle and Astrophysical Xenon (PandaX) experiment\cite{cao2014liquid,zhang2019dark} and the Large Underground Xenon (LUX) experiment \cite{akerib2017results}, which use liquid xenon in detecting dark matter candidates as well as searching for neutrino-less nuclear double-beta decay signals.
	
	Liquid has the advantages of a high transparency and flowability, and is easily purified and filled into the detectors, which makes it a good candidate for the design of particle and nuclear physics experiments\cite{juyal2020proportional,huang2020simulation}, particularly for largescale experiments searching for rare signals\cite{yan2020study} with optical processes and requiring low background environments\cite{giboni2020ln2}.
	
	With liquid-based detectors designed to be increasingly larger, as a growing problem, it is becoming more difficult to keep the uniformity of the liquid throughout an extremely large liquid detector. Moreover, owing to the flowability of the liquid, the liquid flow in the detector may change the uniformity of the optical properties of the liquid. For example, the change in refractive index owing to the temperature and pressure difference may lead to a deviation of the optical photon transport in the medium from ideal uniform conditions. The impact of such deviation caused by a detector with a non-uniformity of the physical signal measurements also needs to be studied, particularly in large-scale experiments searching for rare signal events.
	
	To study the liquid flow and non-uniformity in the detector, it is necessary to have a technique to exchange the detector geometry information between different software programs, such as computational fluid dynamics (CFD) software and detector simulation software. A uniform detector description with a detailed geometry at different parts of the liquid detector needs to be provided, and such information needs to be shared between different software with a common interface. Because the geometry information is dynamic owing to a liquid flow, the transformation of the detector description with the interface should be automatically realized.
	
	We propose the development of a method based on the geometry description markup language (GDML)\cite{chytracek2006geometry} and tessellated detector description\cite{agostinelli2003geant4,allison2006geant4} for sharing the dynamic geometry of liquid-based detectors between CFD simulation software and detector simulation software, making it possible to further study the extent of the nonuniformity of the geometry in the CFD and study its impact on the detector performance.
	
	The rest of this paper is structured as follows. In Sect. 2, we introduce the method of sharing the detector geometry between software and the workflow. In Sect. 3, the method is applied to a toy spherical detector model to study the nonuniformity of the refractive index caused by a liquid flow and its influence on the detector simulation and event reconstruction. In Sects. 4 and 5, the performance of the method and its further application are discussed. Finally, a summary and some concluding remarks are provided in Sect. 6.
	
	\section{Methodologies}
	
	In this section, we describe the method of sharing the detector geometry information between different types of software. With a consistent detector description and automatic geometry data conversion through an interface, the change in detector geometry can be studied in CFD simulation software, and its influence on the detector performance can also be considered using detector simulation software and reconstruction software.
	
	A typical application of the method is in an experiment using optical photons to reconstruct the event vertex and energy of the physical signals, in which the density and refractive index of the liquid medium may change in different parts of the detector owing to changes in the temperature and pressure. Consequently, transportation of the optical photons in the medium will also change, leading to deviations in the detector simulation, event reconstruction, and detector performance.
	
\begin{figure}
		\includegraphics[width=0.8\hsize]{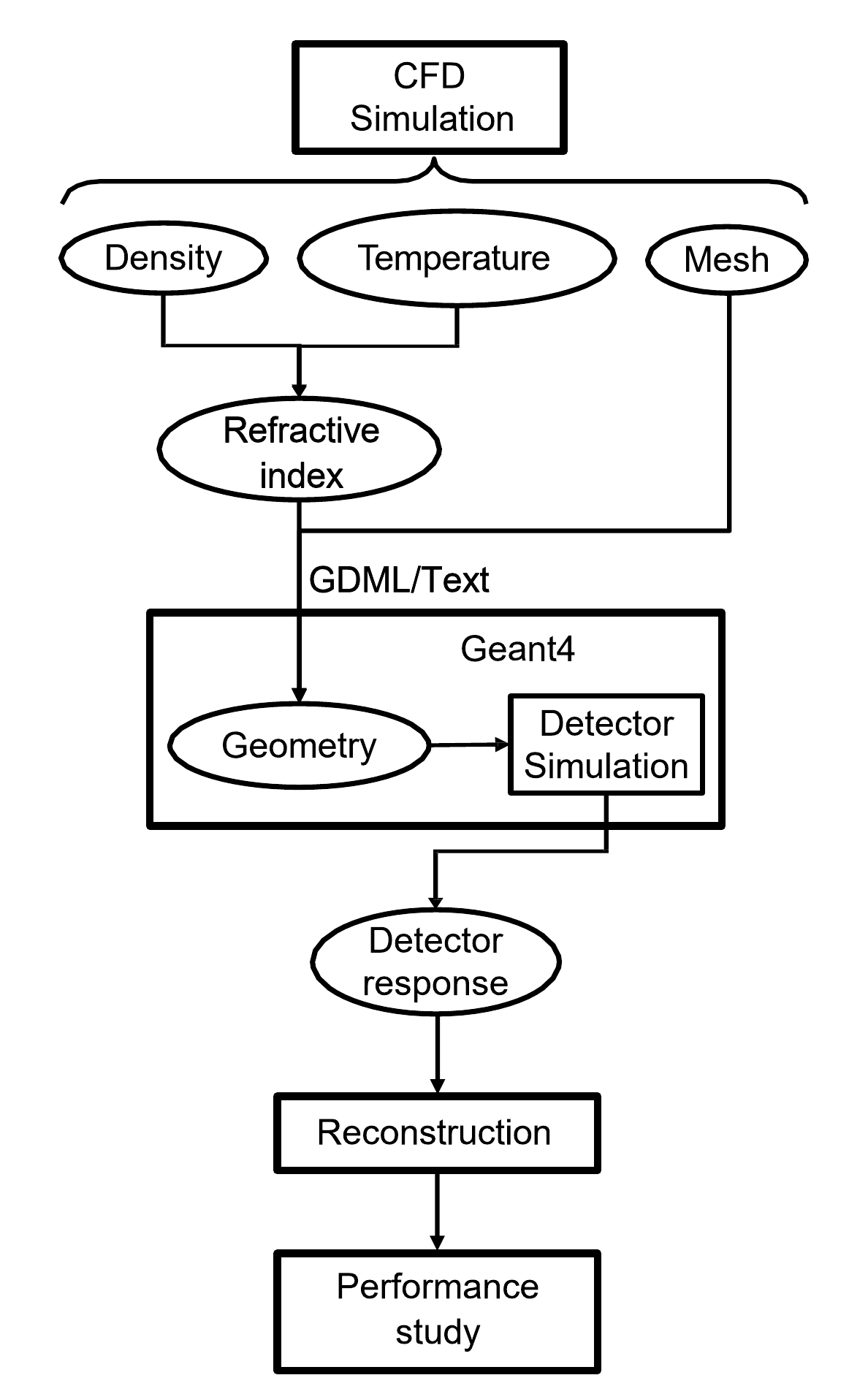}
		\caption{Data flow of the detector geometry information from simulation to reconstruction.}
		\label{data_flow}
	\end{figure}

	The structure of the detector geometry data flow used in the simulation and a reconstruction applying the method is shown in Fig.~\ref{data_flow}. The entire process can be divided into the following parts: a CFD simulation, geometry data conversion, detector simulation, reconstruction, and a detector performance study.
	
	In this method, a CFD simulation is first conducted to study the flow of fluids in the detector, with the detector geometry model constructed and its initial conditions set. After a CFD simulation, the information of the geometry mesh and fluid properties (temperature, pressure, and density distribution) are exported. Meanwhile, the related physics properties of the fluid and the refractive index distribution can also be calculated and exported. The conversion of geometric information from the CFD to the detector simulation is realized using GDML or the text format interface, including the geometric mesh and fluid property information. Next, a geometry with non-uniform medium properties can be built in the detector simulation software, and the propagation process of the particles in the detector with a nonuniform geometry can be simulated. Finally, by comparing the reconstruction results of the uniform detector simulation, the influence of the geometry change on the detector performance is evaluated.
	
	\subsection{CFD simulation}
	
	CFD uses numerical analysis and data structures to analyze and solve problems related to fluid flows. COMSOL Multiphysics\cite{multiphysics1998introduction} and Open-source Field Operation And Manipulation (OpenFOAM) \cite{jasak2007openfoam} are two popular CFD simulation software programs that have been widely used in scientific computations.
	
	COMSOL uses a finite element analysis to decompose objects into a series of small-volume grids for calculation and approximate real-world physical phenomena, particularly to solve the problem of multiphysics field coupling. In our study, the detector is first modeled and meshed, the appropriate fluid and thermal fields are then constructed, and the initial boundary conditions are set. After a finite element analysis and evolution over time, the temperature, density, and flow velocity distribution in each part of the detector become stable and a steady state can be obtained.
	
\begin{figure}[!htb]
		\includegraphics[width=1\hsize]{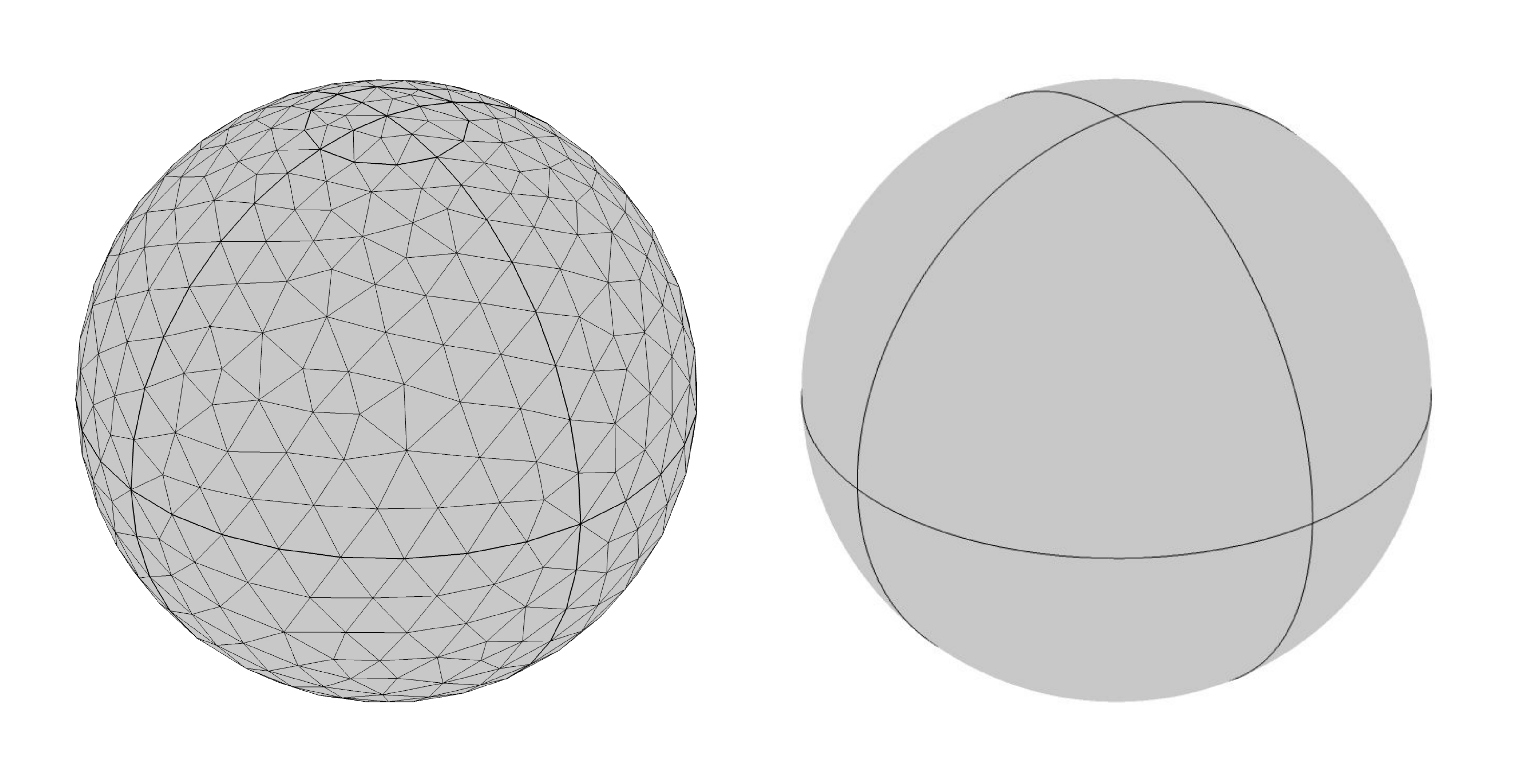}
		\caption{Mesh geometry (left) and CSG shape (right) of a sphere.}
		\label{meshing}
	\end{figure}

	Constructing a mesh of the geometric model is one of the most important steps in a CFD simulation and later studies. The geometric object grid directly determines the solution method of the model and affects the calculation of the problem, including the solution time, the amount of memory required, the interpolation method between the solution nodes, and the accuracy of the solution. The automatic mesh construction function provided by COMSOL can simplify this process; otherwise, a customized mesh needs to be built manually. Fig.~\ref{meshing} shows the description of a sphere with a mesh and a constructive solid geometry (CSG).
	
	OpenFOAMis another popular CFD simulation software. By default, OpenFOAM defines a mesh of arbitrary polyhedral cells in 3D, bounded by arbitrary polygonal faces. The cells can have an unlimited number of faces, and there is no limit to the number of edges or any restriction on its alignment. This type of mesh offers significant freedom in a mesh generation and manipulation, particularly if the geometry of the domain is complex or changes over time. Users can also generate meshes using other packages and convert them into a format that OpenFOAM can use.
	
	\subsection{Geometry data conversion}
	In COMSOL, to construct a mesh of the geometry, the entire detector is divided into many tetrahedrons with a certain precision. The fluid property of the center point of each tetrahedron is taken as the nature of the liquid in the entire tetrahedral area. The mesh and fluid property information of each tetrahedron, including temperature, density, and refractive index, must be converted into a format that can be automatically imported into detector simulation software, such as Geant4\cite{agostinelli2003geant4}, to construct the detector geometry. 
	
	In Geant4, the shape is implemented as a classG4Tessellated Solid, which can be used to generate a generic solid defined by several facets. With such a definition, a complex geometrical shape bounded with surfaces can be converted into an approximate description with facets of a defined dimension. Two types of facets can be used for the construction of a G4TessellatedSolid: a triangular facet and a quadrangular facet. The meshed geometry can be constructed by defining every face of the tetrahedron. Finally, all tessellated solids form the shape of a complete detector.
	
	There are two different ways to realize this process. One is to first generate a GDML file based on the geometric information exported by COMSOL, and then import the GDML file into Geant4 to construct the geometry. GDML is an XML-based\cite{bray2000extensible} geometry description language used to describe the geometry of detectors in physical simulations, including their positions, rotations, shapes, and materials. It is designed as an application-independent persistent format. As pure XML, GDML can be universally used as a format to interchange the geometry between different applications\cite{wang2020cad}. The GDML geometry can be imported into Geant4 using the GDML parser function to construct the detector in Geant4.
	
	GDML has been successfully applied in many HEP experiments\cite{you2008gdml}, such as JUNO\cite{li2018gdml} and BESIII\cite{liang2009uniform}. A GDML-based geometry management system in JUNO offline software is designed to provide a consistent detector description for different applications\cite{you2018root, zhu2019method}. In BESIII, the detector is described using GDML and then applied in a Geant4-based simulation and ROOT-based reconstruction and visualization\cite{liang2009uniform}.
	
	Another method is to directly read the text format output by COMSOL in a Geant4-based simulation code. A complete set of geometry information-sharing interfaces, from COMSOL to Geant4, has been developed. By using volume classes to construct solids for each tetrahedron according to its vertex coordinates and using the G4Material class to customize the material properties of each tetrahedron, the detector geometry with a non-uniform medium is constructed in Geant4.
	
	\subsection{Detector simulation}
	Geant4 is the most popular detector simulation software used in particle and nuclear physics experiments. The simulation process in Geant4 includes the detector construction, particle transport, and interaction with the materials. For a single detector component made up of a uniform material, it is usually described as a single solid volume represented by a CSG shape\cite{dong2020study}. However, for liquid-based detectors, such an approximation will lead to deviation from reality. The level of deviation owing to non-uniformity in large-scale detectors can be too large to be ignored, and thus the difference between a detailed detector description and an approximate uniform description needs to be studied.
	
\begin{figure}[!htb]
		\includegraphics[width=0.9\hsize]{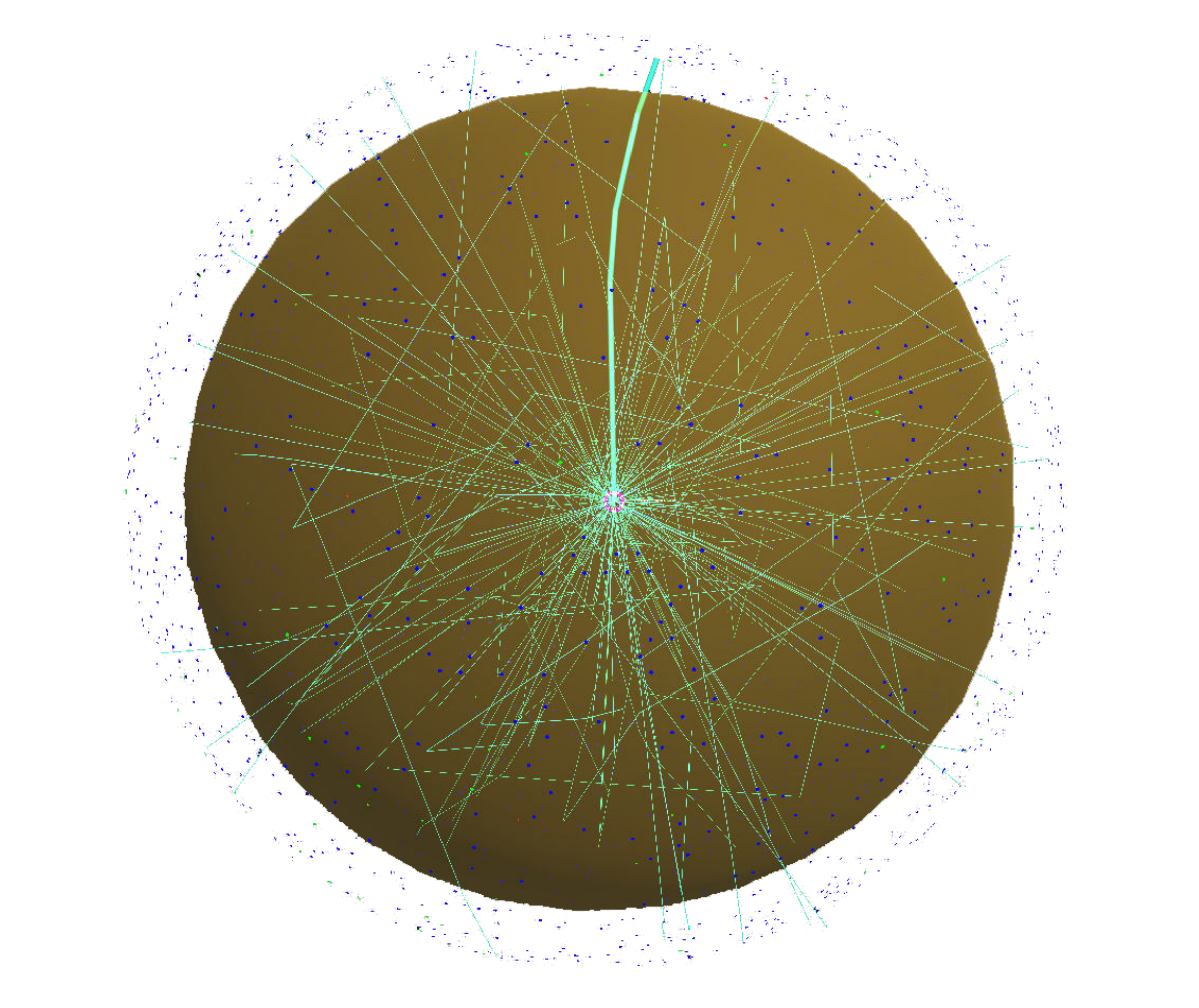}
		\caption{(Color figure online) Schematic of photon transportation in a spherical detector. The bold green line shows the path of a photon with multiple refractions in a non-uniform medium.} 
		\label{optical_path}
	\end{figure}

	To compare the impact of non-uniformity of a refractive index on a detector simulation, only two types of physical processes are included in the detector simulation: photon propagation in a uniform medium and Fresnel refraction at the medium boundary. In the former case, the photon is regarded as a free particle and travels in a straight line throughout the detector. For the latter case, owing to the nonuniform nature of the medium in the detector, the detector has many interfaces. The refractive index on each side of the interface is different, and thus the photon refracts when passing through these interfaces, and the photon path in the detector is composed of a series of steps. As an example, Fig.~\ref{optical_path} shows the refraction of photons in a non-uniform medium.
	
	\subsection{Detector performance study}
	With the output from the detector simulation, the next step is to study the impact of the difference in detector description on the detector performance. The detector performance, for example, the resolution of a certain physical measurement, can be determined by analyzing the detector response during the simulation. The description of the detector geometry plays a key role in the simulation. In this study, we test different descriptions of the detector geometry for their detector performance while leaving the other conditions unchanged.

	\section{Application}
	The idea of the detector geometry sharing method was originally conceived to study the optical photon transportation in liquid-based detectors. Such processes are important in many large-scale liquid-based detector experiments, particularly those searching for rare signal events. For example, in neutrino experiments such as the Daya Bay reactor neutrino experiment\cite{an2013improved}, JUNO, the Kamioka Liquid Scintillator anti-neutrino Detector (KamLAND)\cite{abe2010production}, Hyper-K, and DUNE, optical photons from Cherenkov radiation or scintillation are used to reconstruct the signal events. In a dark matter search or double beta-decay experiments such as PandaX, LUX, Dark matter Experiment with Argon and Pulseshape discrimination 3600 (DEAP-3600)\cite{amaudruz2019design},XENON dark matter search (XENON)\cite{aprile2019light}, and Depleted Argon cryogenic Scintillation and Ionization Detection (DarkSide)\cite{agnes2018low}, photons or electrons are also transported in liquid detectors.
	
	In this section, we describe the use of a toy detector model to demonstrate the feasibility of applying the method in a photon transportation study. A simple spherical volume filled with water is constructed as the detector. Assuming that the temperature distribution is not uniform in the spherical detector, the density of the water in the detector will be different, which makes the water flow. This step is calculated using CFD simulation software COMSOL with a meshed geometry. The refractive index of the water is determined based on the temperature and difference in density. The fluid properties and mesh geometry information are then automatically converted to set up the geometry in Geant4 with tessellated solids.
	
	In a Geant4-based detector simulation, a virtual physical event is generated at a fixed vertex inside the sphere, and multiple photons are emitted from the vertex. The photons propagate in the water detector and finally reach the surface of the sphere. The detector simulation outputs are the hit positions of the photons on the spherical surface, which will be used to reconstruct the position of the event vertex.
	
	To study how a non-uniformity of the liquid can affect the detector performance, the detector is constructed using two methods. One is to construct a single non-uniform water sphere, and the other is to construct the same water sphere with a tessellated geometry, with the fluid property of each tetrahedron set from the CFD simulation result. The simulation and reconstruction output will be different by using the two detector descriptions, from which the deviation and difference in performance can be studied.
	
	\subsection{Detector mesh geometry}
\begin{figure}
		\includegraphics[width=1\hsize]{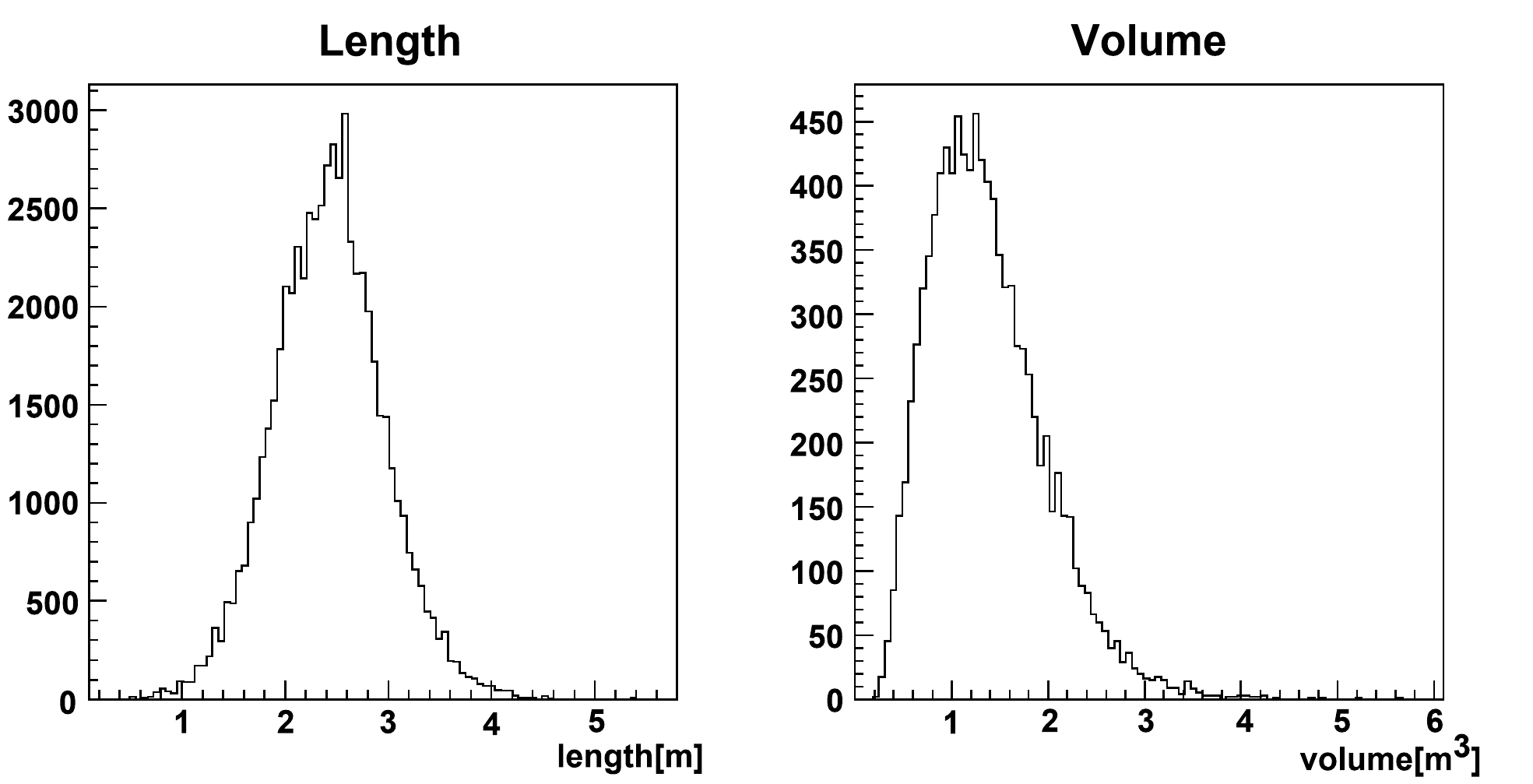}
		\caption{Distribution of the tetrahedron side length and volume size in a meshed sphere. 
		}
		\label{mesh}
	\end{figure}

	In the CFD simulation, the radius of the sphere is set to 15 m, and the mesh is configured to divide the sphere evenly into approximately 10,000 tetrahedrons. The distribution of the side length and volume size of every tetrahedron is shown in Fig.~\ref{mesh}, respectively. No tetrahedron has a particularly long side length or particularly large volume, which guarantees that the detector is evenly segmented; otherwise, the properties of the medium in the central tetrahedral area cannot be used to approximate the properties of the liquid within the whole tetrahedral area.
	
\begin{figure}
		\centering
		\subfigure{
			\begin{minipage}[t]{0.5\hsize}
				\centering
				\includegraphics[width=\hsize]{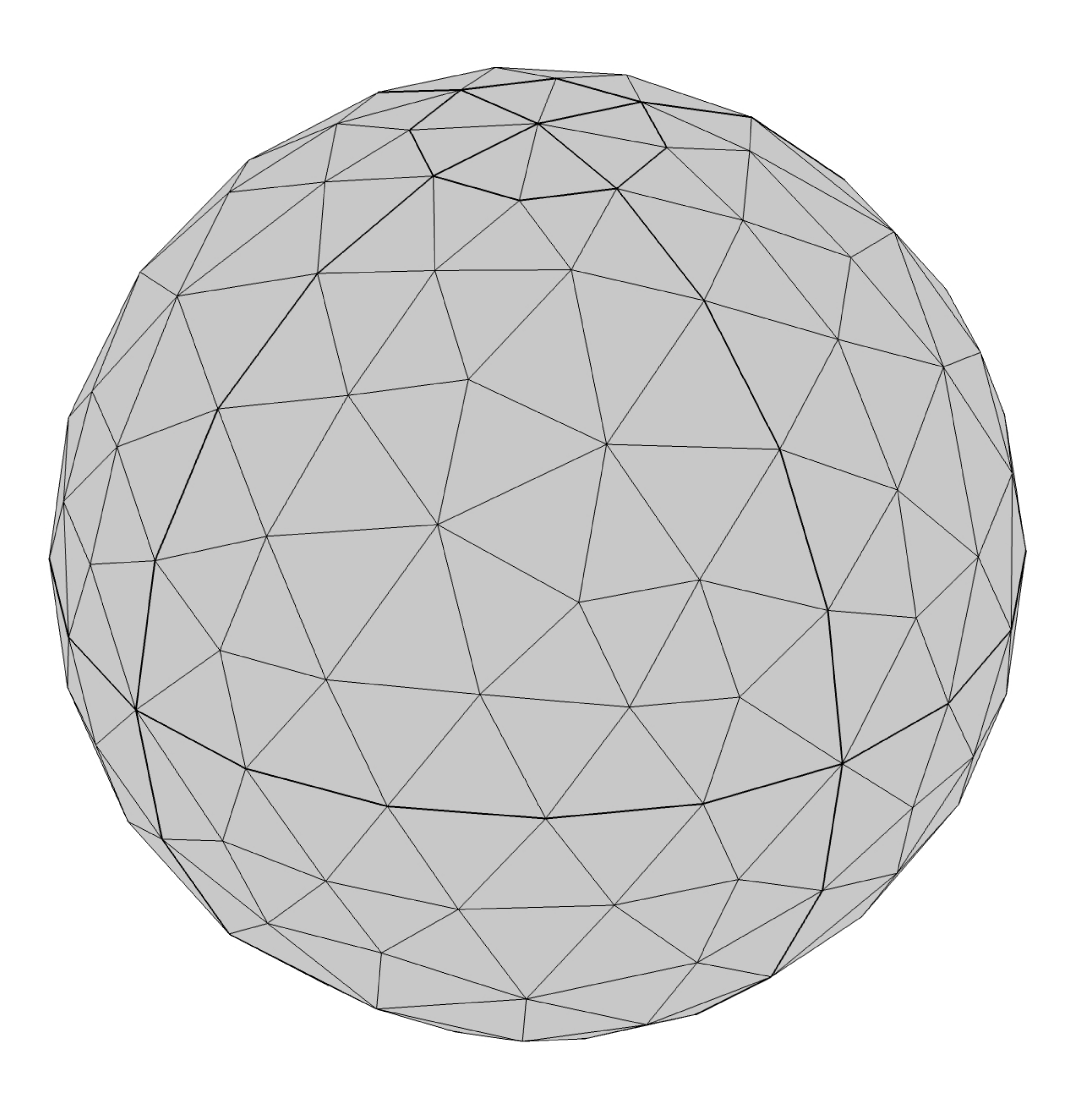}
			\end{minipage}
		}%
		\subfigure{
			\begin{minipage}[t]{0.5\hsize}
				\centering
				\includegraphics[width=\hsize]{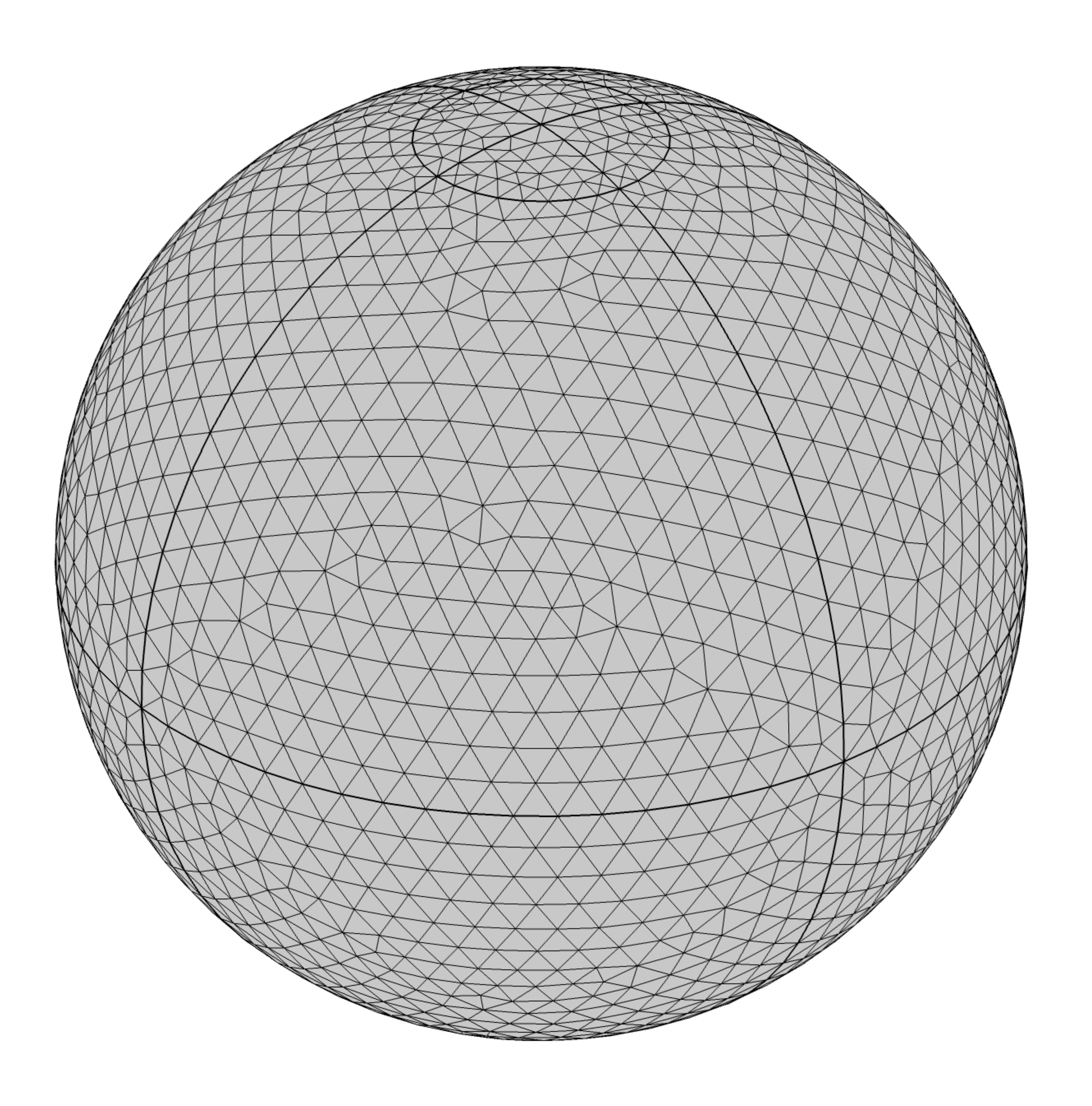}
			\end{minipage}
		}%
		\caption{Mesh geometry of the sphere with different granularities of 1000 tetrahedrons (left) and 100,000 tetrahedrons (right).}
		\label{mesh_diff_tet}
	\end{figure}

	The precision of the detector description depends on the granularity of the tessellated geometry. The more tetrahedrons the sphere is divided into, the more precise the detector is described, and the more computing resources are required in the CFD and detector simulations. Fig.~\ref{mesh_diff_tet} shows the division of the sphere into 1000 tetrahedrons and 100,000 tetrahedrons. In Sect. 4, the dependence of the deviation of the vertex reconstruction on the number of tetrahedrons is studied. The correlations between the program running speed and the number of tetrahedrons are also provided.
	
	\subsection{Fluid properties in CFD simulation}
	
	In the CFD simulation based on COMSOL, a fixed temperature difference is set between the top and bottom of the detector as the boundary conditions. To make the difference caused by temperature more significant and to test the feasibility of the method, a temperature difference of 35$^{\circ}$C is set in a preliminary study. The correlation of the performance difference with temperature will be provided in Sect. 4.

	The temperature at the top is set to higher than that at the bottom. A constant heat flux is provided in the top and bottom areas to simulate the temperature difference from the environment.

\begin{figure}[!htb]
		\includegraphics[width=\hsize]{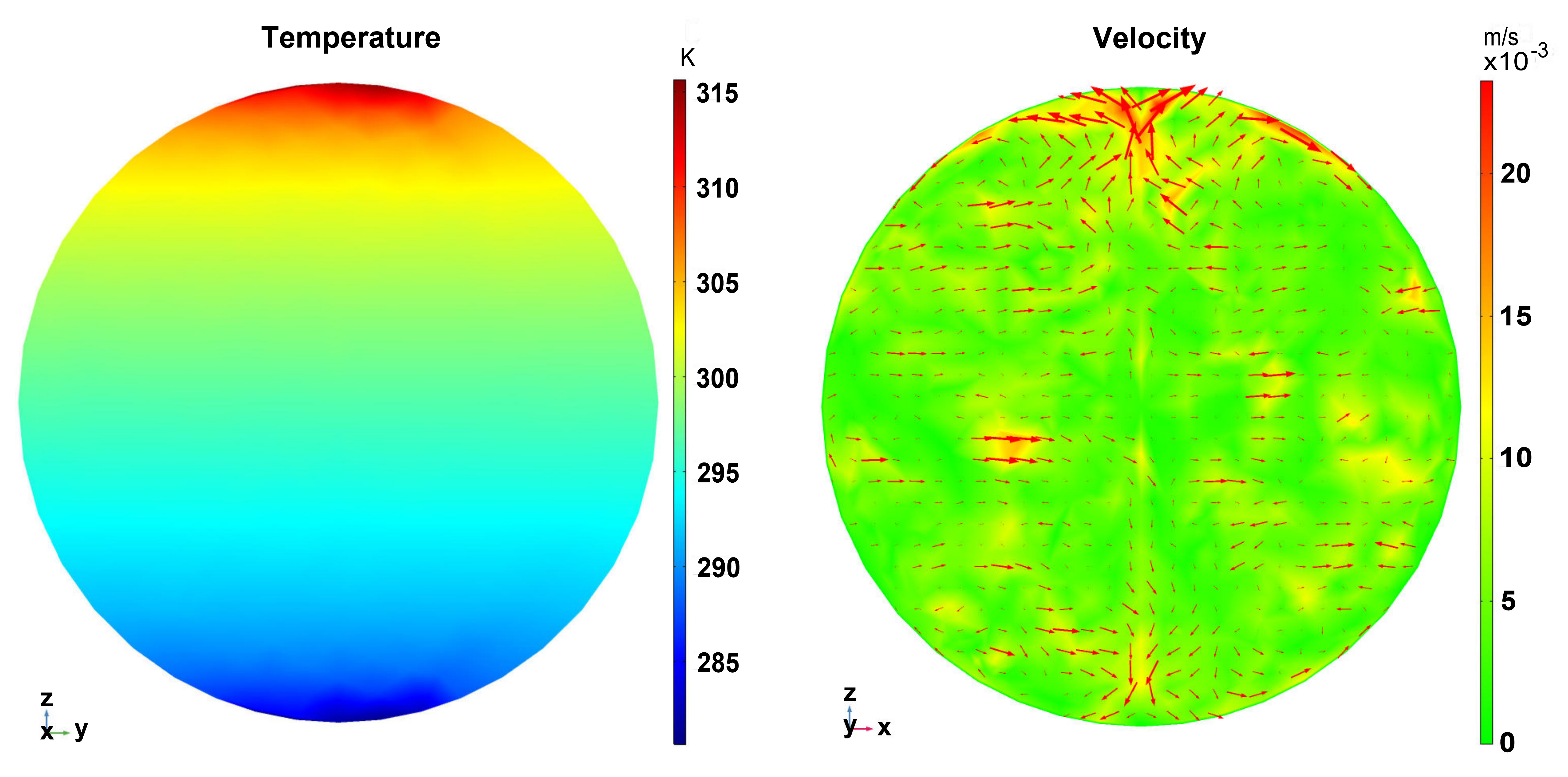}
		\caption{Fluid temperature (left) and flow velocity (right) distribution in the spherical detector. }
		\label{simulation_sphere}
	\end{figure}

	The simulation results show that when there is a temperature difference between the upper and lower surfaces of the detector, a stable temperature gradient is formed in the medium. Fig.~\ref{simulation_sphere} shows the distribution of fluid temperature and flow velocity at a steady state. The temperature gradient direction is along the Z-axis, and the temperature is approximately the same on the X-Y plane. The medium in the sphere only slowly flows horizontally on the X-Y plane, except for the area near the north and south poles.
	
	The refractive index of a liquid is highly dependent on the temperature and density. According to\cite{schiebener1990refractive}, the refractive index of water in a different grid of the mesh can be calculated using the following equation:

	\begin{equation}
	\begin{aligned}
	\frac{n^2-1}{(n^2+2)\bar\rho} &= a_0 + a_1\bar\rho + a_2\bar T + a_3{\bar \lambda}^2\bar T + a_4/{\bar \lambda}^2 \\ &+ \frac{a_5}{{\bar \lambda}^2-{{\bar \lambda}_{UV}}^2} + \frac{a_6}{{\bar \lambda}^2-{{\bar \lambda}_{IR}}^2} + a_7{\bar\rho}^2
	\\
	\bar\rho &= \rho/\rho^* , \bar T = T/T^* , \bar \lambda = \lambda/\lambda^*
	\end{aligned}
	\end{equation}
	
	where $n$ is the refractive index of the liquid, $\rho$ is the density of the liquid, $T$ is the temperature of the liquid, and $\lambda$ is the wavelength of the photon. The other parameters used in the formula are listed in Table ~\ref{parameters}.

	\begin{table}[!htb]
		\center
		\caption{Parameters of the refractive index of water with wavelength, density, and temperature}
		\setlength{\tabcolsep}{3mm}{%
			\begin{tabular}{@{}lll@{}}
				\toprule
				\midrule
				$a_0=0.244257733$ &   & $a_4 = 1.58920570\cdot 10^{-3}$\\
				$a_1=9.74634476\cdot 10^{-3}$ &   & $a_5=2.45934259\cdot 10^{-3}$ \\
				$a_2=-3.73234996\cdot 10^{-3}$ &   & $a_6=0.900704920$ \\
				$a_3=2.68678472\cdot 10^{-4}$ &   & $a_7=-1.66626219\cdot 10^{-2}$ \\
				${\bar \lambda}_{UV}=0.2292020$ &   & ${\bar \lambda}_{IR}=5.432937$ \\
				$\rho^*=1000 \rm kg\cdot m^{-3}$ &   & $T^*=273.15 \rm K$ \\
				$\lambda^*=0.589 \rm \mu m$ &   & \\
				\bottomrule        
		\end{tabular}}%
		\label{parameters}
	\end{table}
	
	\subsection{Deviation in detector simulation}
	
	The mesh of the spherical detector and its fluid properties in different grids of the mesh, including the density and refractive index, will be converted into a GDML format and text format that can be read automatically by the Geant4 simulation program to construct the detector geometry.

	To simulate a physics event at a fixed vertex inside the sphere, 1000 optical photons are generated from the vertex isotropically at the $4\pi$ solid angle. Physically, the number of photons generated corresponds to the amplitude of the physical event, such as the incident particle energy. In the Geant4 simulation, the optical photons are transported in the water sphere through optical processes and finally reach the surface of the sphere. Ideally, we assume that all photons can be recorded as hits at the sphere surface with 100\% detection efficiency. These signals will be used to reconstruct the original vertex of the physics event.
	
	To obtain the simulation deviation, we first simulate using a uniform medium detector to obtain the nominal simulation outputs. Then,we run a second simulation with all parameters unchanged, except for a replacement of a uniform sphere with a tessellated geometry describing a non-uniform medium.

\begin{figure}[!htb]
		\includegraphics[width=1\hsize]{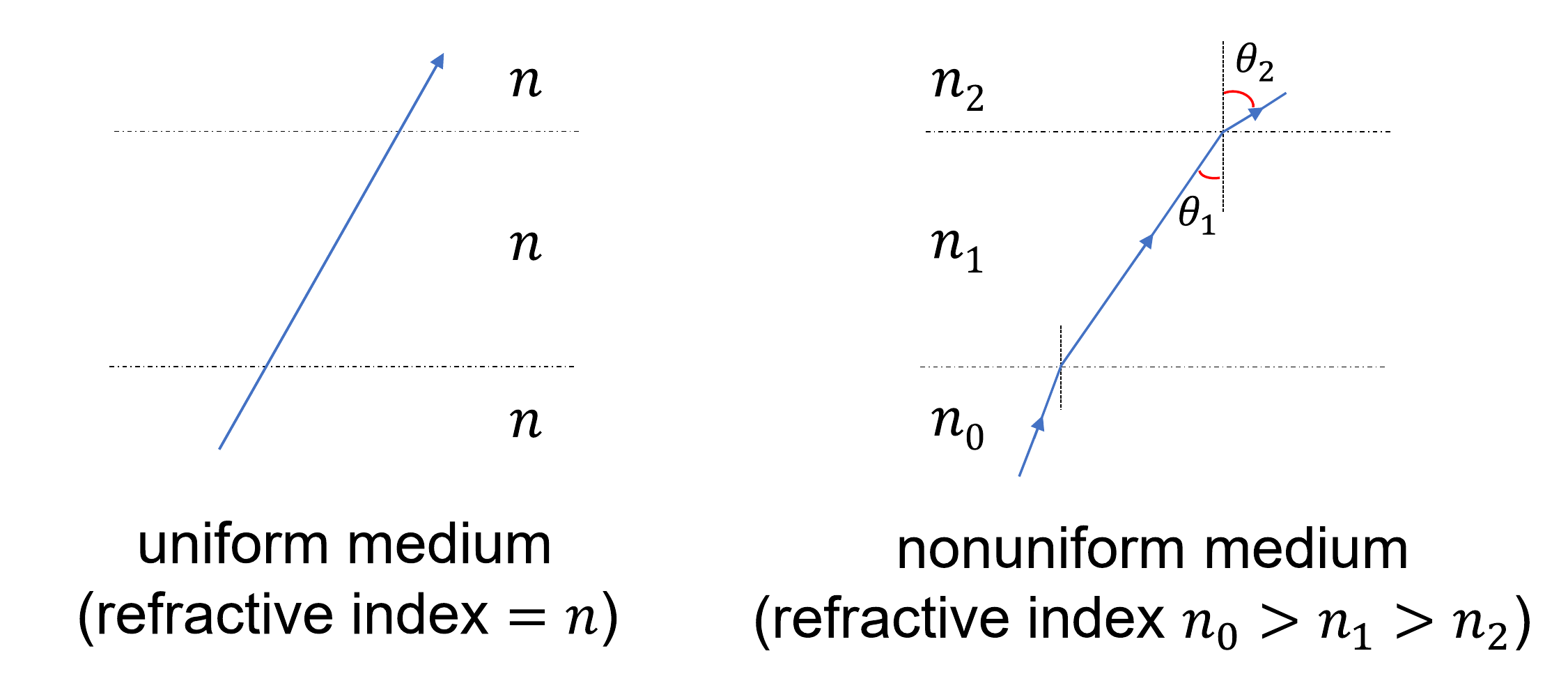}
		\caption{Two scenarios with refraction. The left shows that the photon travels in a straight line in the detector described with a uniform medium. The right shows that the photon refracts multiple times in the detector with a non-uniform medium.}
		\label{refraction}
	\end{figure}

	Fig.~\ref{refraction} describes the scenario using the above two geometries in a detector simulation. One uses a uniform material in the detector construction, and the other uses a tessellated geometry for a more precise detector description. For the optical photons crossing the neighboring tessellated volumes with different refractive indices, only the Fresnel refraction is processed.
	
	The magnitude of the photon momentum at the sphere surface is the same as the incident momentum, but the direction of the photon changes in each Fresnel refraction, which is described by the following equations:
	
	\begin{equation}
	\begin{aligned}
	\boldsymbol{p}_2 = \boldsymbol{p}_1 + | \boldsymbol{p}_1|(\cos\theta_1-\frac{n_2}{n_1}\cos\theta_2)\hat{\boldsymbol{n}}
	\end{aligned}
	\end{equation}
	\begin{equation}
	\begin{aligned}
	\cos\theta_1 = \frac{|{\boldsymbol{p}_1} \cdot {\hat{\boldsymbol{n}}}|}{|\boldsymbol{p}_1|} , \sin\theta_2 = \frac{n_1}{n_2}\sin\theta_1
	\end{aligned}
	\end{equation}
	
	where $n_1$ and $n_2$ are the refractive indices on different sides of the interface, respectively. 
	$\theta_1$ is the incident angle, and $\theta_2$ is the refraction angle. 
	$\boldsymbol{p}_1$ is the incident momentum, $\boldsymbol{p}_2$ is the momentum after refraction, and $\hat{\boldsymbol{n}}$ is the normalized normal vector of the refraction plane.
	
	To obtain the relation of simulation deviation with the event vertex, 10,000 events are simulated with vertices randomly distributed in the sphere. Each event is simulated twice with two different detector descriptions to obtain the deviation. We can then compare the photon hit positions on the detector surface owing to the refraction caused by the nonuniformity of the medium.

\begin{figure}[!htb]
		\centering
		\subfigure{
			\includegraphics[width=1\hsize]{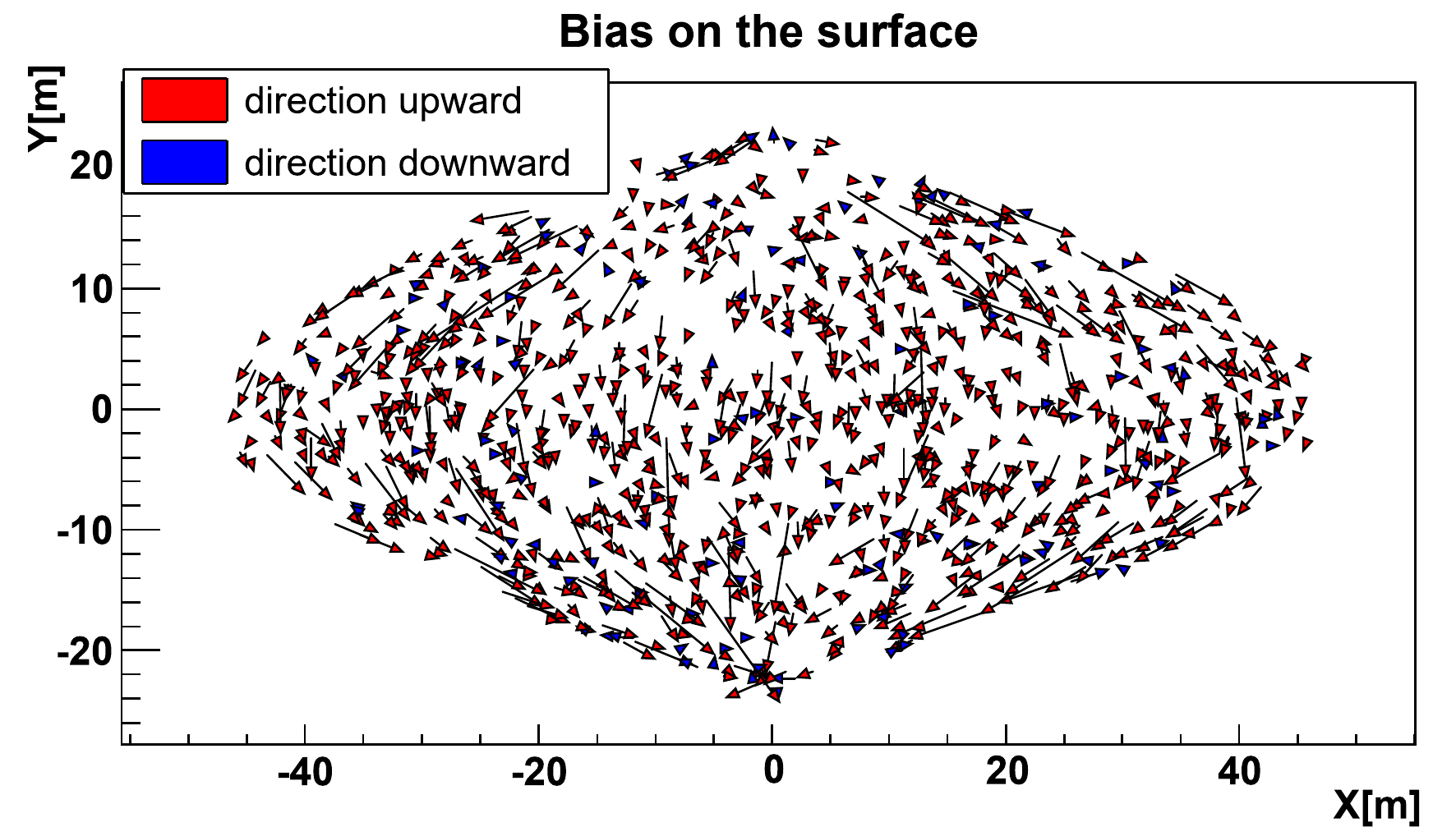}
		}
		\quad
		\subfigure{
			\includegraphics[width=1\hsize]{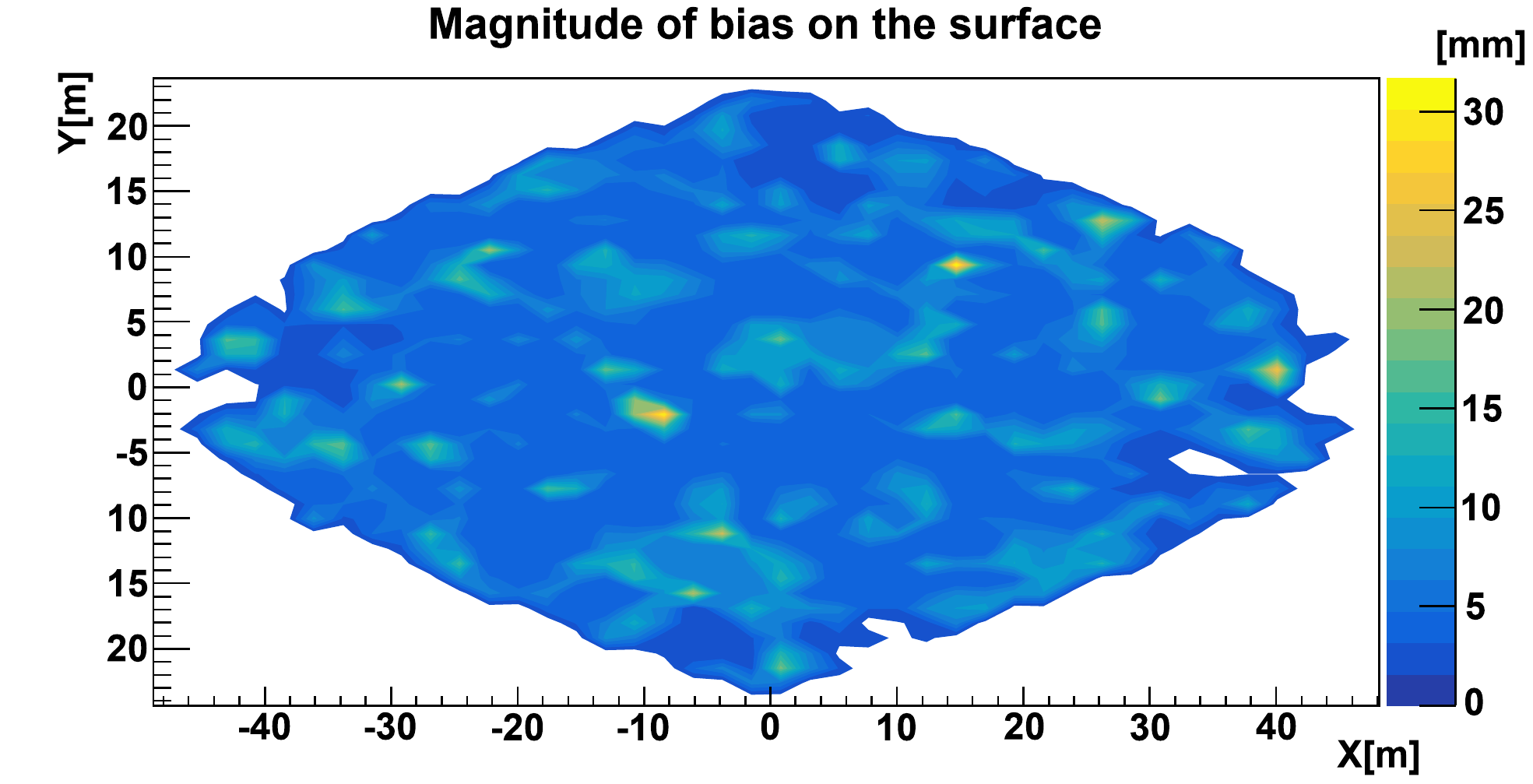}
		}
		\caption{(Color figure online) Distribution of the magnitude and direction of the deviation on the detector surface (top). 
			The blue and red arrows indicate the offset direction along the positive Z-axis and negative Z-axis directions, respectively. 
			The length of the arrow indicates the relative magnitude of the deviation.
			A 2D projection of the magnitude of the offset on the sphere surface is also shown (bottom).}
		\label{arrow_and_proj}
	\end{figure}

	A comparison of the results from two simulations shows that the hit position on the detector surface is significantly shifted under the preliminary conditions. The deviation is at the millimeter level. Fig.~\ref{arrow_and_proj} shows the distribution of the magnitude and direction of the deviation on the surface of the sphere. The magnitude of deviation has no obvious relationship with the hit position on the sphere. Most of the photon positions on the surface of the sphere deflect toward the negative Z-axis in the simulation with the non-uniform detector, regardless of the original upward or downward direction of the photon from the vertex points. This is because the temperature increases along the positive Z-axis direction and the refractive index decreases along this direction, as shown in Fig.~\ref{simulation_sphere}.

\begin{figure}[!htb]
		\centering
		\subfigure{
			\includegraphics[width=1\hsize]{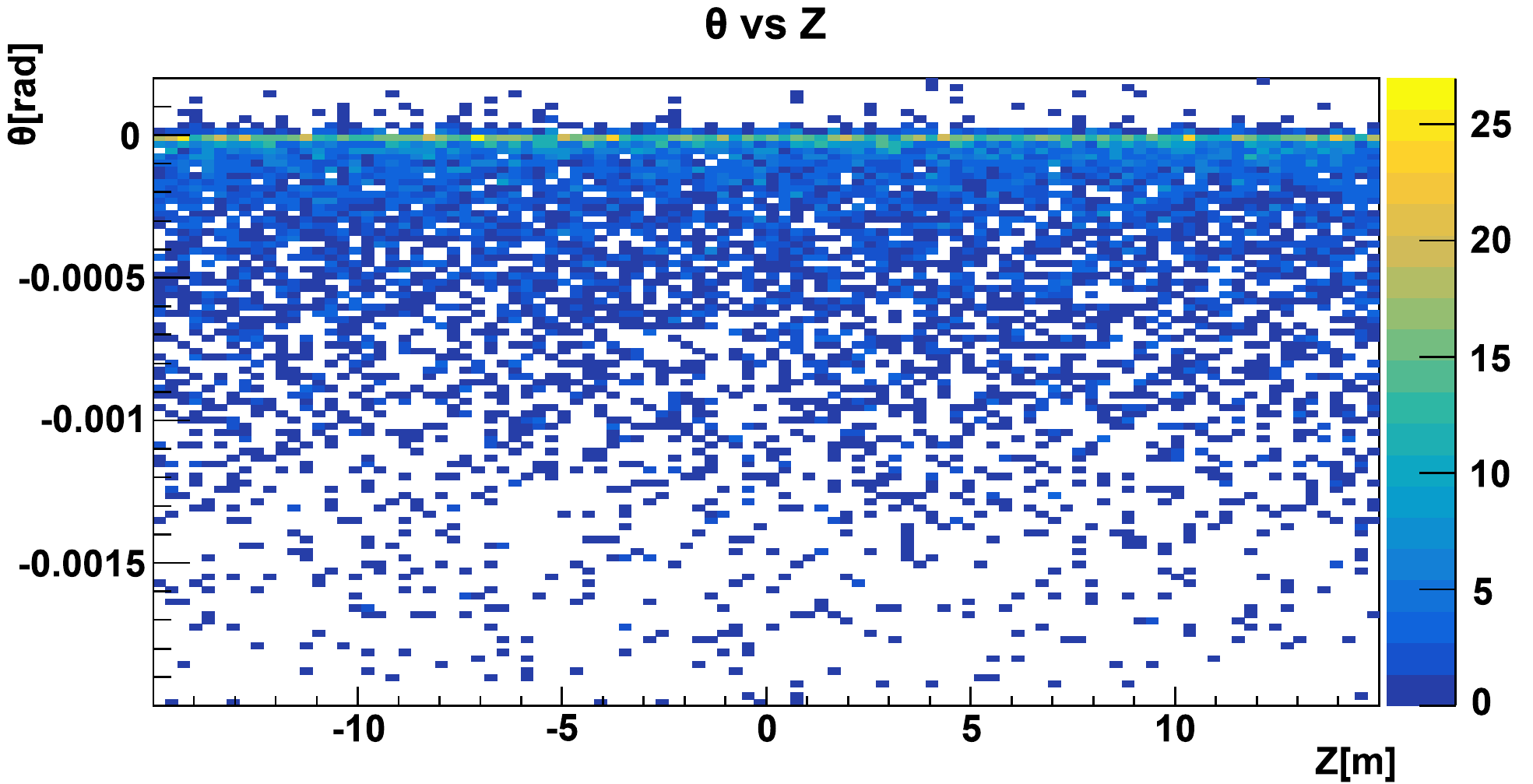}
		}
		\quad
		\subfigure{
			\includegraphics[width=1\hsize]{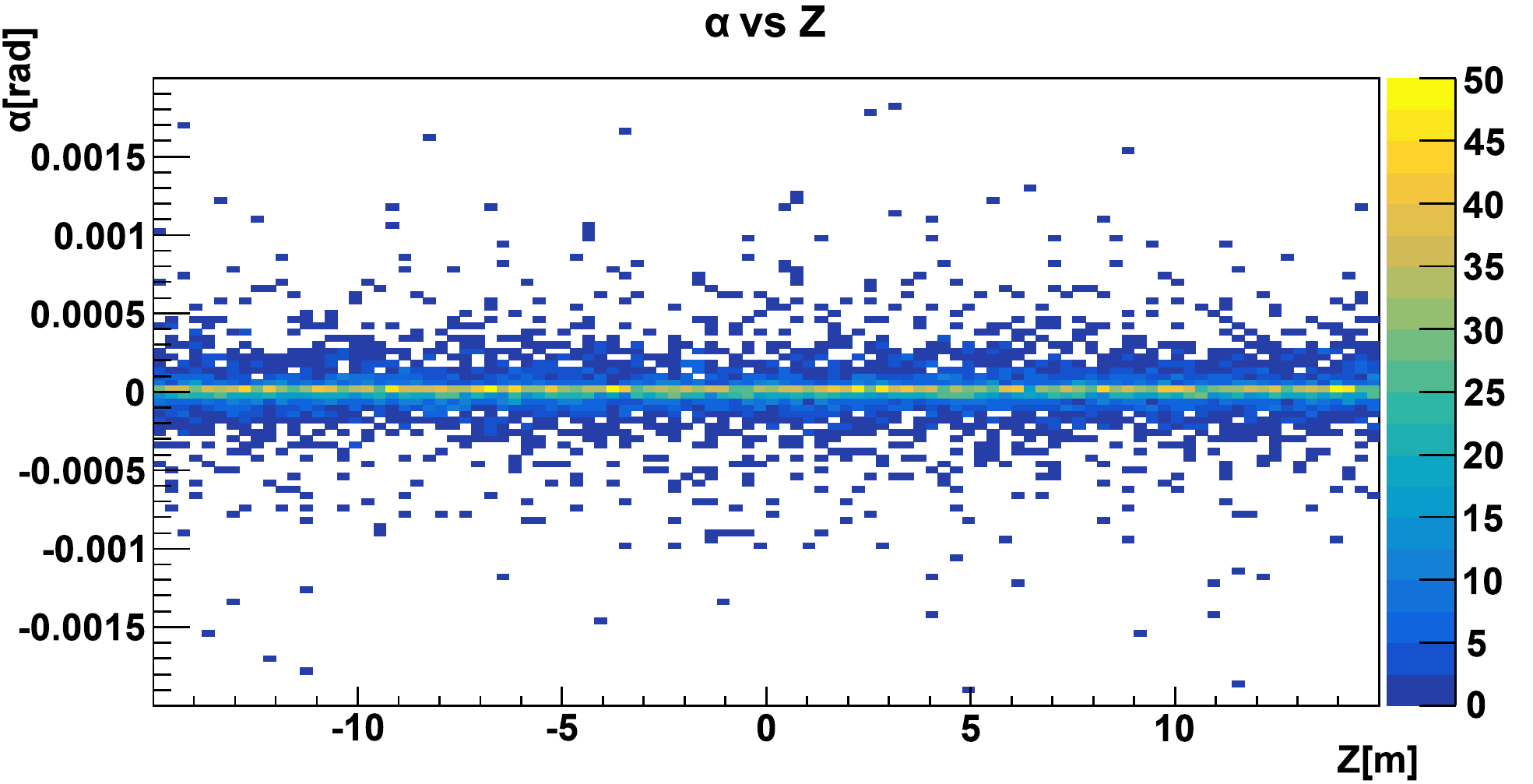}
		}
		\caption{Distribution of the deviation angle $\theta$ (top) and $\alpha$ (bottom) with the Z-coordinate.}
		\label{theta_and_alpha}
	\end{figure}

\begin{figure}[!htb]
		\includegraphics[width=0.6\hsize]{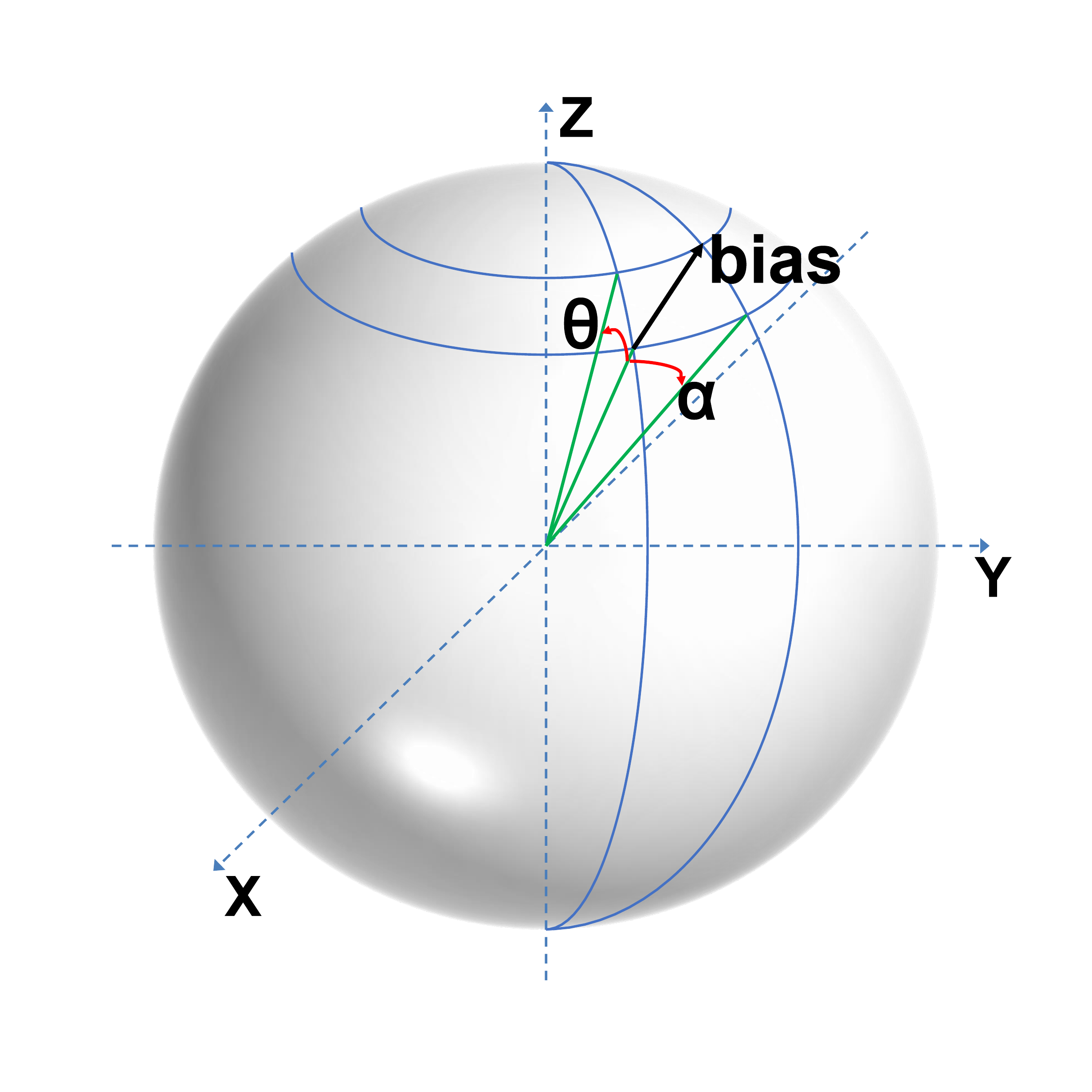}
		\caption{Schematic view of the deviation angle on sphere surface. Here, $\theta$ and $\alpha$ represent the deviation angle component along the longitude and latitude directions, respectively.}
		\label{angle}
	\end{figure}

	Fig.~\ref{theta_and_alpha} shows the distribution of the deviation angle offset in the longitude ($\theta$) and latitude ($\alpha$) directions, which are defined in the schematic view of the sphere in Fig.~\ref{angle}. Because the temperature on the same $Z$-plane is the same, there is no significant trend in which the photon hits with the same $Z$ on the surface deflect systematically toward the left or right.
	
	\subsection{Deviation in vertex reconstruction}
	
	In most experiments, the event vertex position, which shows where the physics event is generated, is an important physics quantity to measure. The resolution of the vertex position is a key indicator for evaluating the detector performance. To study the effect of the non-uniformity of the medium properties on the resolution of the detector vertex position, a charge-weighted algorithm is used to reconstruct the event vertex from the hit position of the photons on the sphere surface.

\begin{figure}[!htb]
		\includegraphics[width=0.6\hsize]{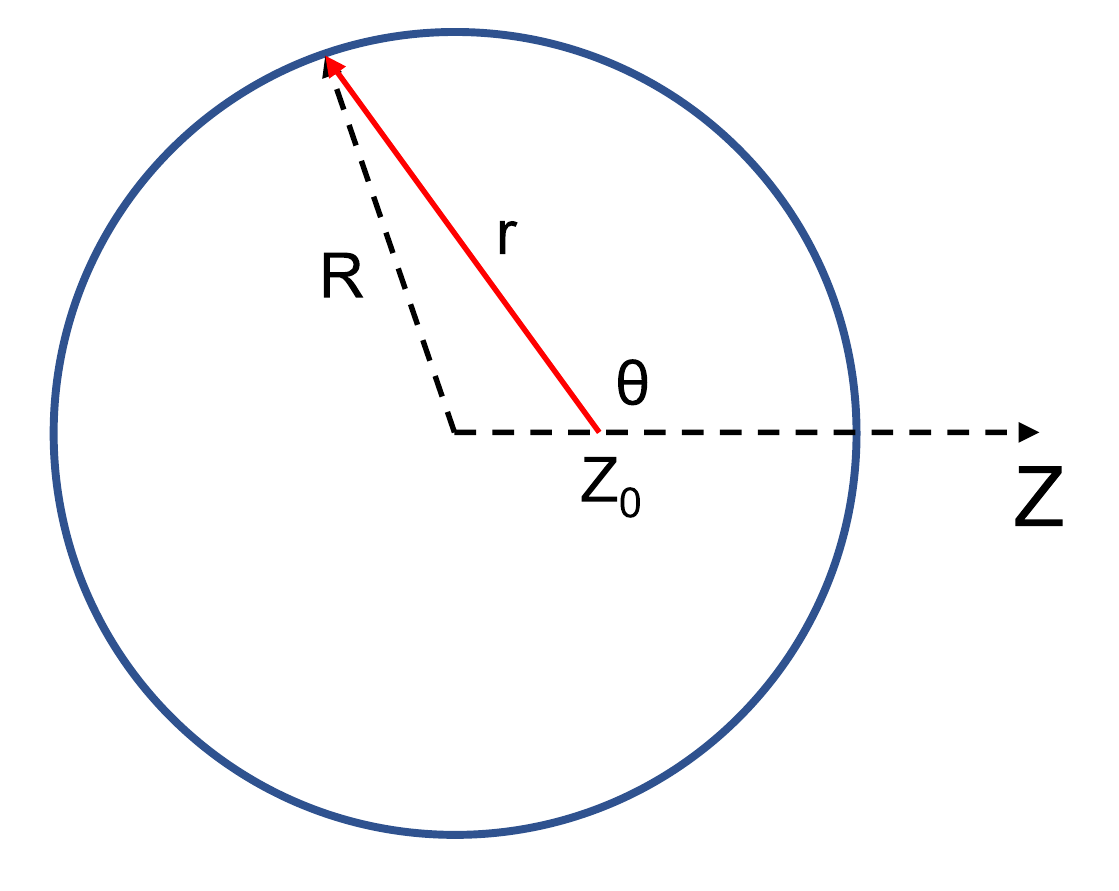}
		\caption{Sketch of the charge-weighted algorithm for vertex reconstruction.}
		\label{cc_principle}
	\end{figure}

	Fig.~\ref{cc_principle} shows the principle of the charge-weighted algorithm. Assuming that the true event vertex is located at the position $Z=z_0$ and that there are $n$ photons emitted from the vertex, whose momentum directions are isotropically distributed in the $4\pi$ solid angle of the vertex, the average value of the $z$ component of the hit position of all photons on the sphere is
	
	\begin{equation}
	\begin{aligned}
	\bar{z} &= \frac{1}{4\pi}\int z\,d\Omega 
	\\ &= \frac{1}{4\pi}\int_0^{2\pi}\,d\phi \int_0^{\pi} (z_0 + r\cos\theta)\sin\theta\,d\theta
	\\ &= \frac{2}{3}z_0
	\end{aligned}
	\end{equation}
	The reconstructed event vertex $\bar{\boldsymbol{V}}$ can then be obtained as follows:
	\begin{equation}
	\bar{\boldsymbol{V}} = \frac{3}{2n}\sum_i \boldsymbol{V}_i
	\end{equation}
	where $n$ is the total number of photons and $\boldsymbol{V}_i$ is the hit position of the $i$-th photon on the sphere.
	
\begin{figure}[!htb]
		\includegraphics[width=0.8\hsize]{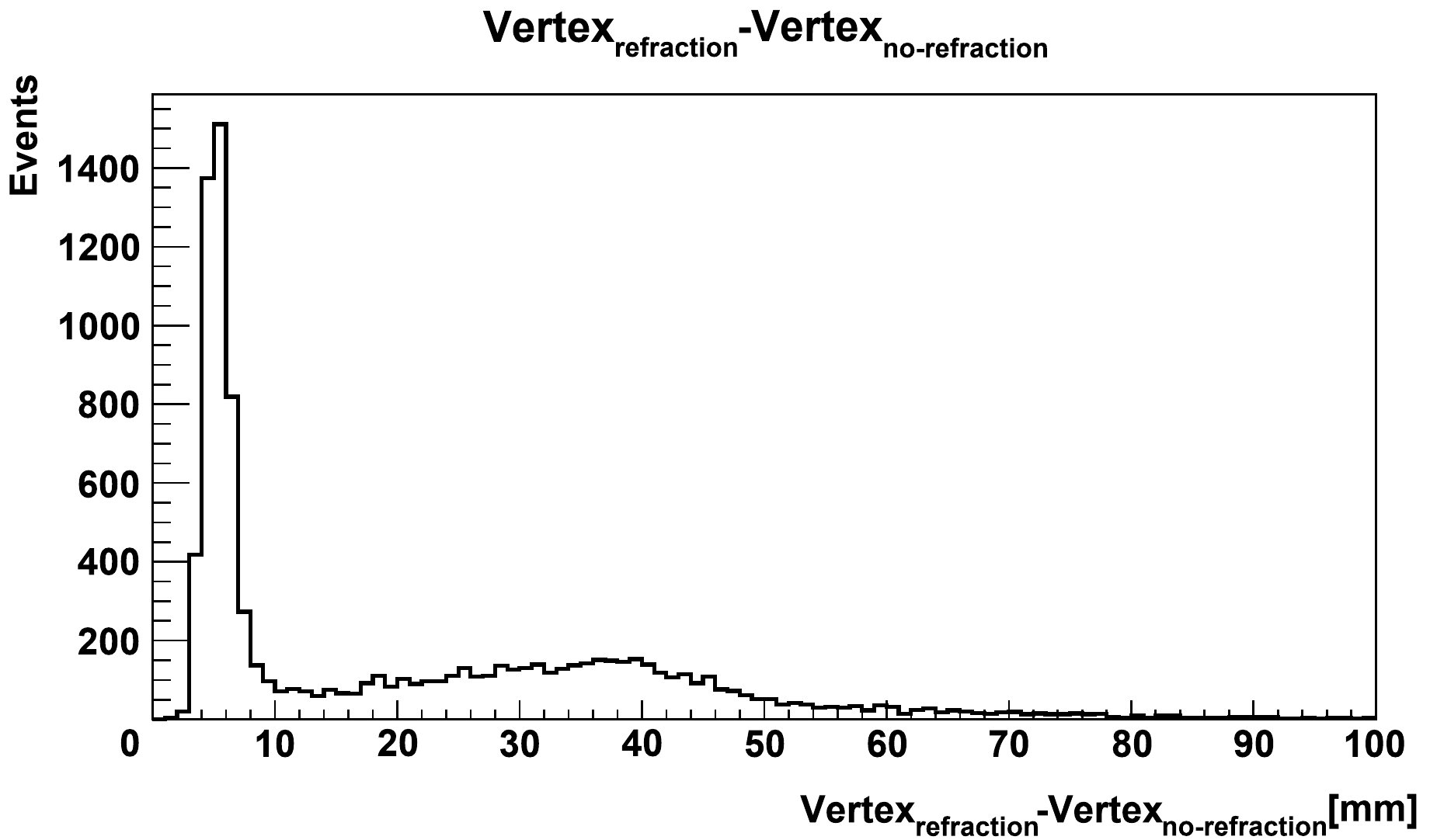}
		\caption{The vertex reconstruction deviation owing to difference in geometry.}
		\label{bias_v}
	\end{figure}

	To compare the event vertex reconstruction deviation, we simulate 10,000 events. Each event was simulated twice, with a uniform and non-uniform detector geometry, respectively. The two simulation results are reconstructed with the same charge-weighted algorithm. We then compare the two reconstructed vertices to obtain the deviation in this event owing to the difference in geometry. Fig.~\ref{bias_v} shows the distribution of reconstructed vertex deviation for the 10,000 events. The deviations owing to the difference in geometry are a few millimeters in length with a long tail of up to 80 mm. The average deviation is approximately 10 mm. Although the charge-weighted algorithm has a vertex reconstruction resolution itself, in the two simulations of every event, every original condition is the same except for the detector geometry. Therefore, the comparison is between the two reconstructed vertices rather than their respective bias from the true vertex.
	
	In this section, we use a set of fixed-parameters (sphere radius = 15 m, temperature difference = 35$^{\circ}$C, and sphere mesh number = 10,000) to obtain preliminary results showing the feasibility of applying the geometry sharing method to a liquid-based detector simulation and reconstruction, and to obtain the deviation distribution.
	
	\section{Comparison}
	
	To study the influence of different conditional parameters on the detector performance, we change the size of the detector, the temperature difference, and the mesh granularity to check the consequential changes in the simulation and vertex reconstruction results.
	
	\begin{figure}[!htb]
		\includegraphics[width=0.8\hsize]{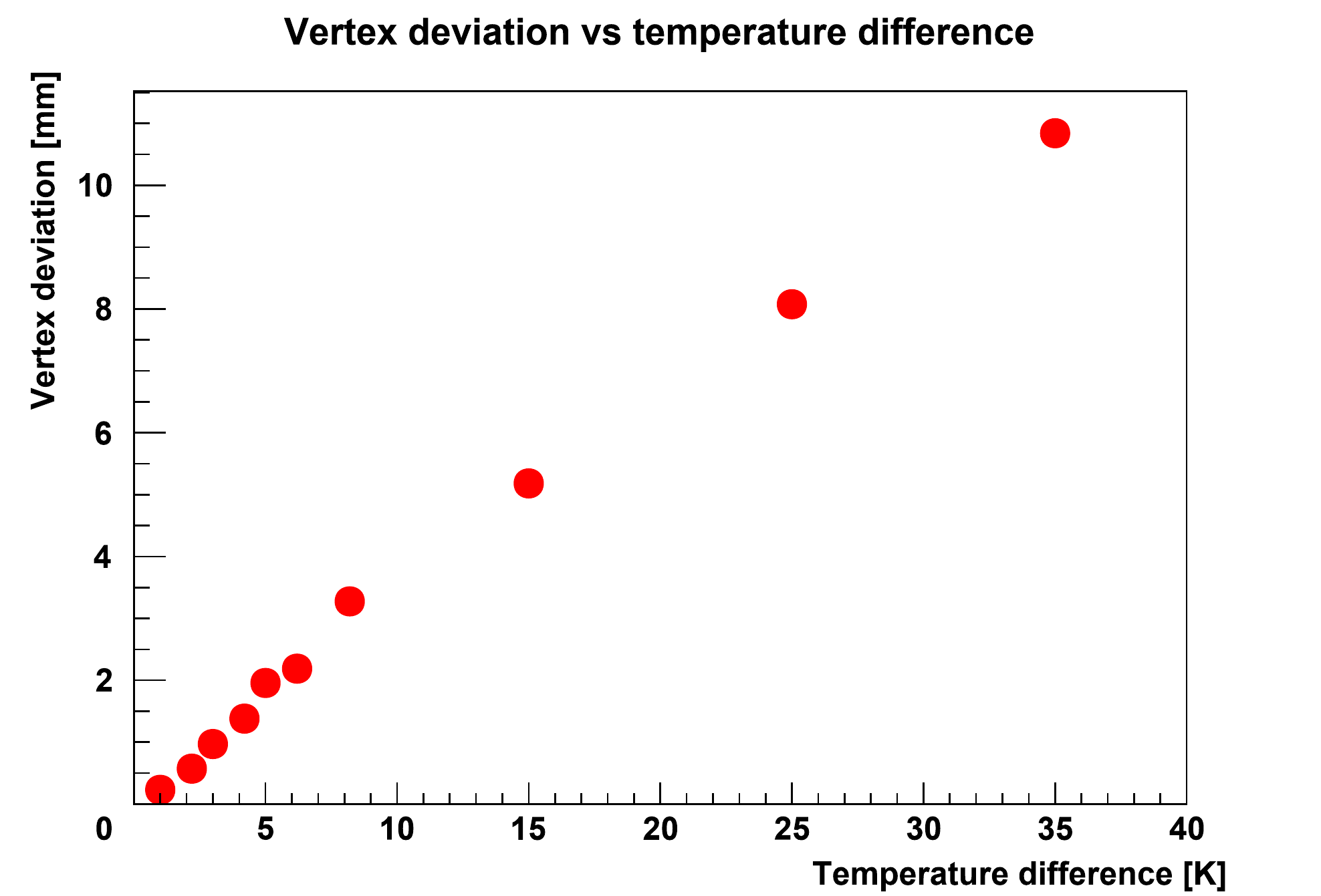}
		\caption{Average deviation of the reconstructed vertex changes with the temperature difference.}
		\label{diff_dT}
	\end{figure}
	\begin{figure}[!htb]
		\includegraphics[width=0.8\hsize]{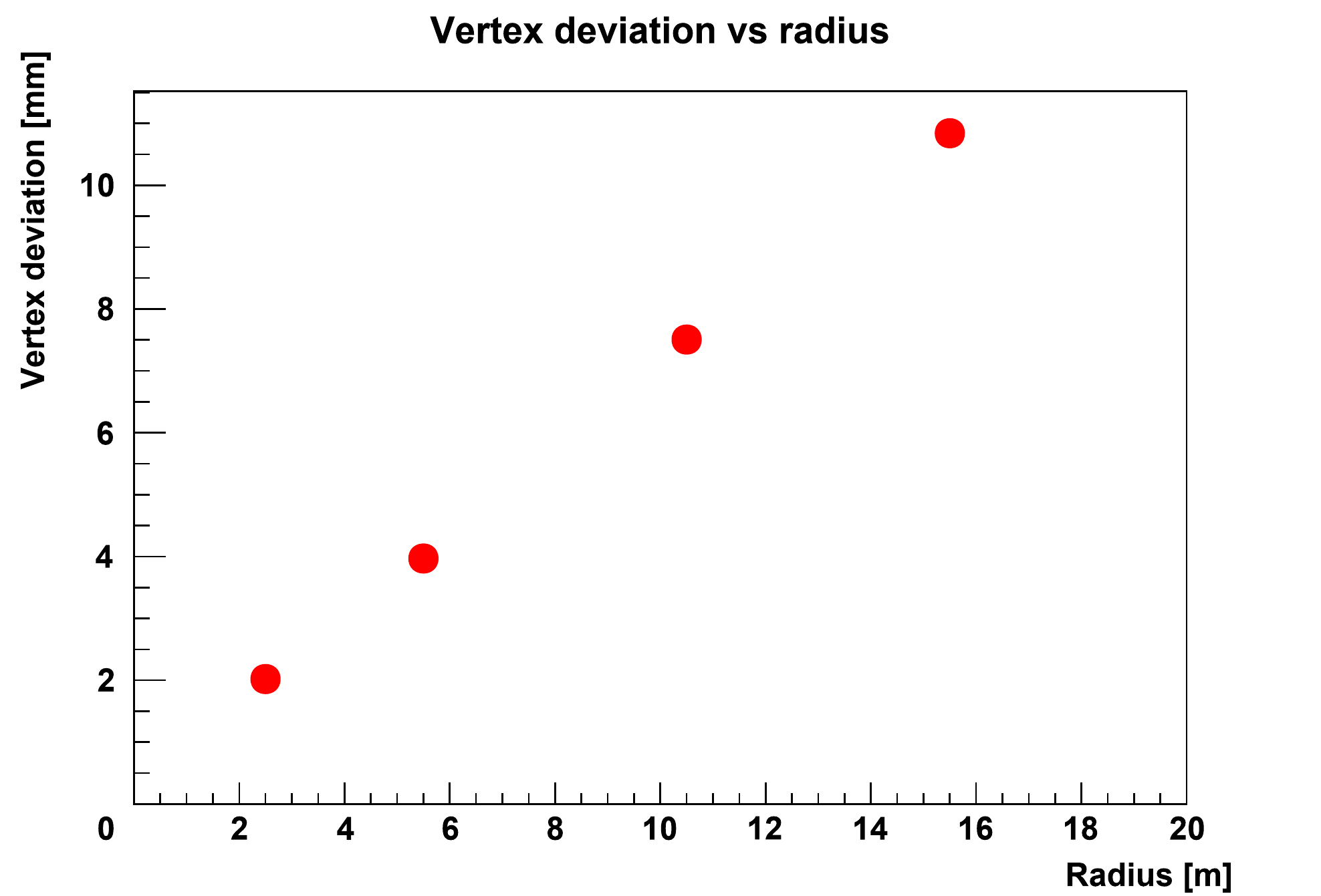}
		\caption{Average deviation of the reconstructed vertex changes with the radius of the sphere.}
		\label{diff_R}
	\end{figure}

	Under real experiment conditions, the environment temperature could be different. Usually, the temperature difference is at a few degrees Celsius. The higher it is, the more significant its impact on the detector performance. Fig.~\ref{diff_dT} shows that the reconstructed vertex deviation is linearly dependent on the temperature difference. Fig.~\ref{diff_R} shows that the reconstructed vertex deviation is linear with the size of the detector. These conclusions are expected because the method itself is size independent.

\begin{figure}[!htb]
		\includegraphics[width=0.8\hsize]{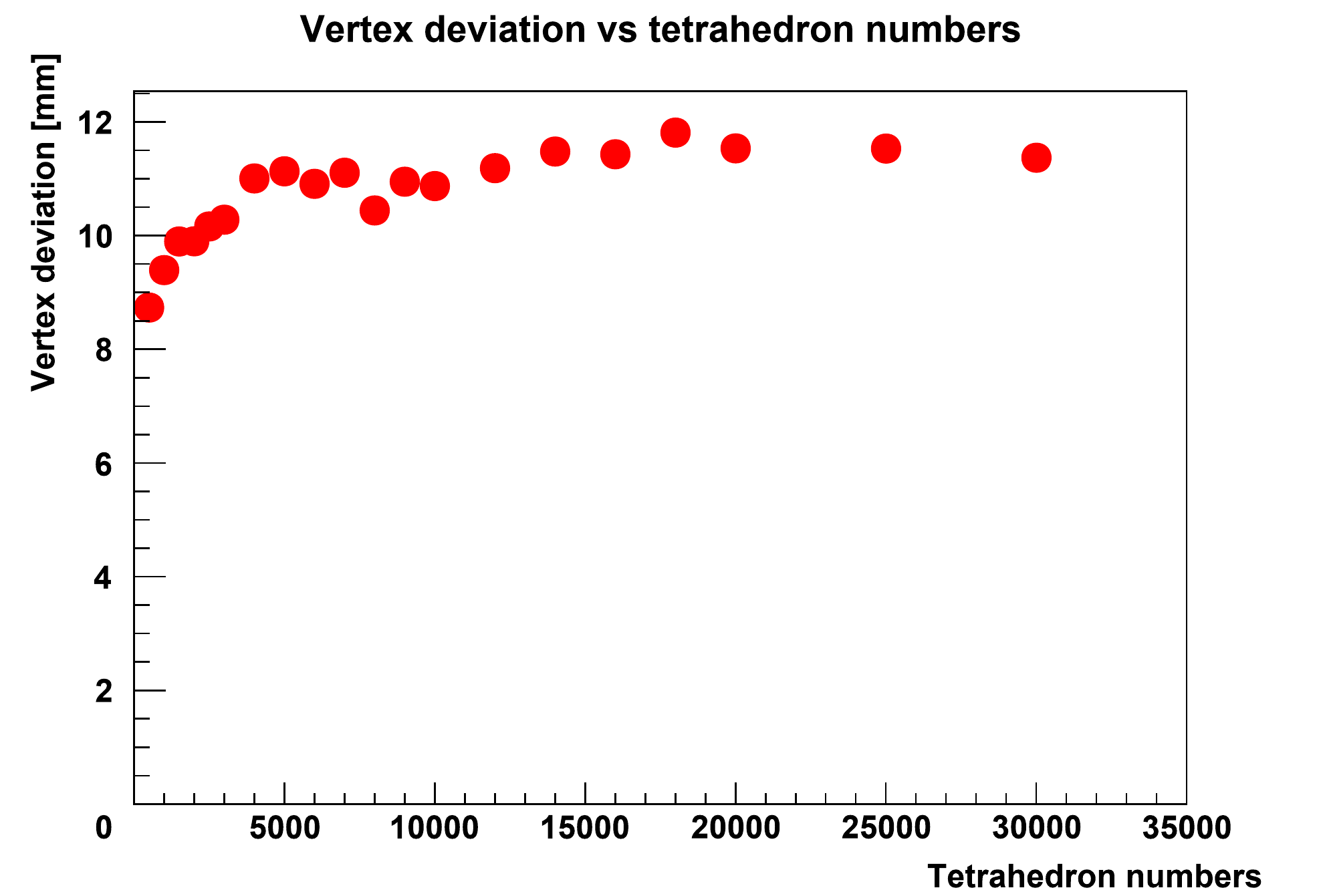}
		\caption{The average deviation of the reconstructed vertex changes with the segmentation accuracy.}
		\label{diff_tetnum}
	\end{figure}

	Fig.~\ref{diff_tetnum} shows the dependency of the vertex reconstruction deviation on the mesh granularity, \emph{i.e.}, the number of tetrahedrons that a sphere is segmented into. When the segmentation is extremely coarse, the deviations of the reconstructed vertex gradually increase with the improvement in the segmentation accuracy. After the number of tetrahedrons reaches 5000, the deviation becomes stable. To determine the granularity of the detector segmentation, it is necessary to comprehensively consider the available computing resources and resulting accuracy requirements.
	
\begin{figure}[!htb]
		\includegraphics[width=0.8\hsize]{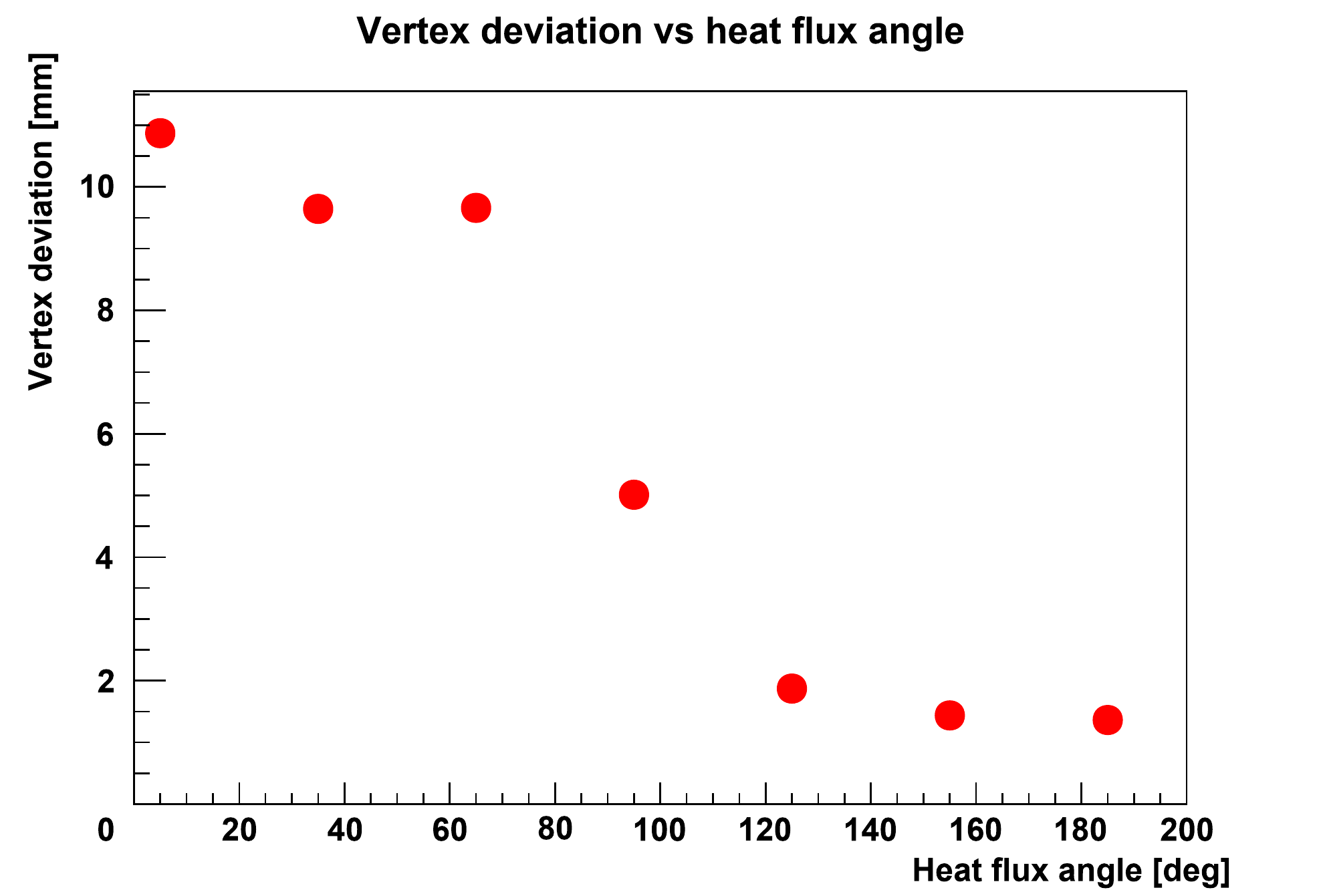}
		\caption{The average deviation of the reconstructed vertex changes with the heat flux area angle.}
		\label{diff_angle}
	\end{figure}

	Fig.~\ref{diff_angle} shows the deviation dependency on the heat flux area angle. When the heat flux area is close to the top or bottom of the sphere, the deviation of the reconstruction vertex is slightly affected by the heat flux location change, but when the location deviates greatly from the $Z$-axis by nearly $90^{\circ}$, the deviation of the reconstructed vertex is significantly reduced.
	
\begin{figure}[!htb]
		\includegraphics[width=1\hsize]{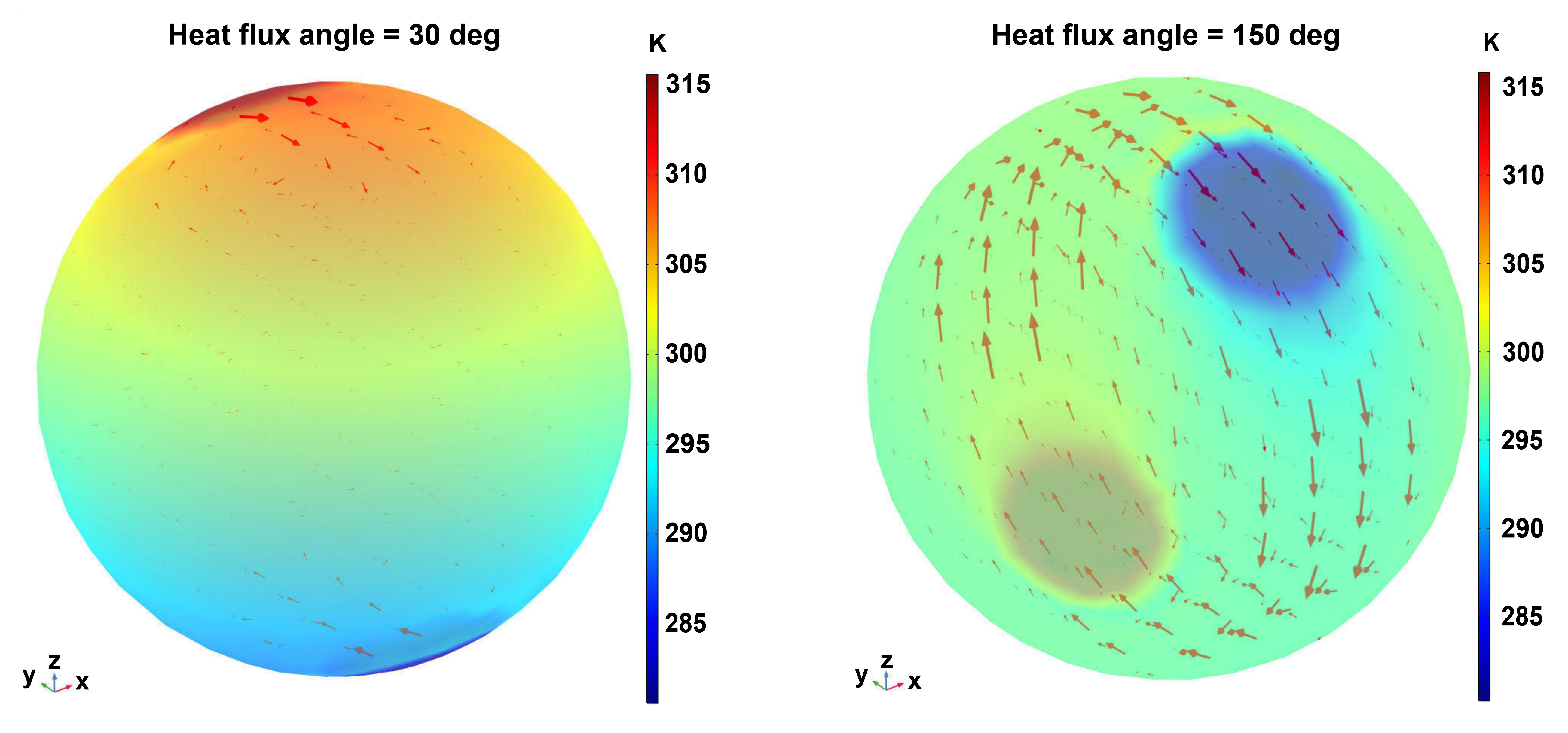}
		\caption{The distributions of temperature and liquid flow in the sphere when the CFD simulation reaches stability for the two cases when the heat flux area deviates from the $Z$-axis by $30^{\circ}$ (left) and by $150^{\circ}$ (right), respectively.}
		\label{diff_angle_v}
	\end{figure}

	As the reason for the influence of the location of the heat flux area, the formation of the temperature gradient has a strong correlation with the location of the heat flux area. Fig.~\ref{diff_angle_v} shows the temperature distribution that reaches stability when the heat flux region deviates from the $Z$-axis by $30^{\circ}$ and $150^{\circ}$, respectively, in which the red and blue areas on the surface represent hot and cold areas where a heat flux exists. When the heat flux area deviates from the $Z$-axis by $30^{\circ}$, the medium heated in the hot area increases along the ball wall to the top and decelerates, and then flows back to the hot area along the horizontal ball wall. A similar case occurs in the cold area. Therefore, the liquid in the sphere has almost no vertical flow, but mainly a horizontal flow.
	
	Because there is only local circulation, when the heat flux area deviates from the $Z$-axis by an angle that is not too large, a temperature gradient similar to the case in which the heat flux area is on the $Z$-axis can still be formed. When the heat flux area deviates far from the $Z$-axis, the temperature difference between the upper and lower endpoints along the temperature gradient direction under a stable state becomes smaller.
	
\begin{figure}[!htb]
		\includegraphics[width=0.9\hsize]{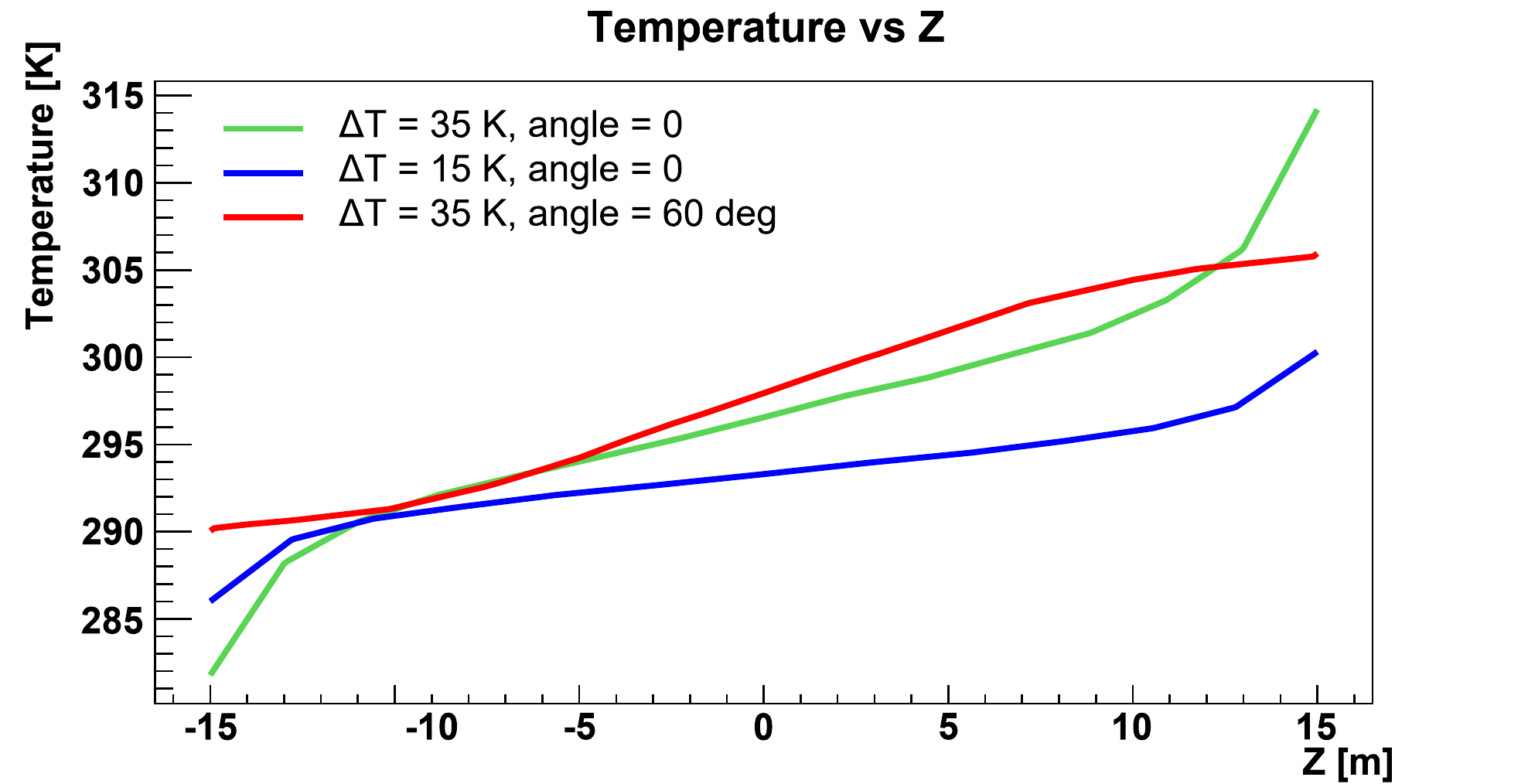}
		\caption{The temperature distribution in the sphere along the $Z$-axis with different ${\Delta}T$ and different heat flux angle.}
		\label{t_vs_z}
	\end{figure}

	Fig.~\ref{t_vs_z} shows the temperature distribution along the $Z$-axis at a steady state. When the temperature difference of the heat flux is 35 K and the heat flux location deviates from the $Z$-axis by $60^{\circ}$, the temperature difference is comparable to the case in which the temperature difference of the heat flux is 15 K and the position is along the $Z$-axis, although the temperature gradient of the former is greater. However, when the heat flux area deviates from the $Z$-axis by more than $90^{\circ}$ (Fig.~\ref{diff_angle_v}, right), the heated liquid flows upward from the hot area, and when it reaches the cold area on the top, the liquid is quickly cooled and sinks. Therefore, it forms an annular flow in the sphere, and no obvious temperature gradient can be formed owing to the circulation.

	The consumption of computing resources is also compared in this study. When the number of tetrahedrons increases with the improvement in the segmentation accuracy, it takes more CPU time in the COMSOL and Geant4 simulations. The simulation times with different numbers of tetrahedrons are listed in Table ~\ref{cpu_time}, which is the CPU time required for COMSOL to simulate a fluid flow for a 5 h clock time and for Geant4 to simulate 100,000 particles. The amount of CPU time used by the two geometry data conversion methods based on the text format and GDML is also compared. In the Geant4 simulation, it takes more time to import the geometry through a GDML file than directly reading the text file to construct the geometry. With an increase in the segmentation accuracy, the difference in the computing speed between the two methods becomes more obvious.

	\begin{table}[!htb]
		\caption{CPU time consumption for each part of the simulation}
		\begin{tabular}{ccccc}
			\hline
			\multirow{3}{*}{tetrahedron numbers} & \multicolumn{3}{c}{Geant4}                                                 & \multirow{3}{*}{COMSOL} \\ \cline{2-4}
			& \multicolumn{2}{l}{Geometry initialization} & \multirow{2}{*}{simulation} &                         \\ \cline{2-3}
			& Text                       & GDML            &                             &                         \\ \hline
			1,000                      & \textless{}1s             & 8s              & 27s                         & 21min                   \\
			10,000                     & 2s                        & 13s             & 58s                         & 137min                  \\
			30,000                     & 7s                        & 44s             & 85s                         & 321min                  \\ \hline
		\end{tabular}
		\label{cpu_time}
	\end{table}
	
	\section{Discussion}
	With this method, the generation of a correct geometry mesh is an important step. The entire detector must be fully segmented; otherwise, the properties of the tetrahedral central medium cannot be used to approximate the properties of the entire tetrahedron. In this study, only the method of a global automatic segmentation is adopted, and only a tetrahedron segmentation is considered.

	At the same time, the influence of medium non-uniformity will reach saturation with an improvement in the segmentation accuracy. An excessively high segmentation accuracy does not significantly improve the simulation and reconstruction precision but will greatly increase the CPU time consumption in the CFD and Geant4 simulations. In more realistic detector simulations, more complicated physical processes, more particles, and more complicated environments are usually introduced, which may further increase the simulation time.

	The GDML and text format are both used for geometry data sharing from COMSOL to Geant4. Creating a geometry by directly reading the text files is much faster than importing GDML files into Geant4 to initialize the geometry. However, GDML is widely used in a Geant4 simulation and reconstruction based on the ROOT\cite{brun1997root} geometry package, which can provide consistent geometric information for a subsequent data analysis.

	In our research, it is assumed that the hot area is on the top and the cold area is at the bottom, and thus a temperature gradient can be formed. When the location of the hot area is lower than that of the cold area because of the formation of convection and circulation in the sphere, the temperature in the sphere is almost the same and no temperature gradient can be formed. The entire spherical detector may be considered to be uniform, and no obvious deviations in simulation and reconstruction will be observed.

	\section{Conclusion}
	
	A method of dynamic geometry information sharing between a CFD simulation and a detector simulation is proposed for liquid-based detectors. Its feasibility is demonstrated by applying the method to a simulation with a non-uniform medium to study the photon transport and a deviation in the event of a vertex reconstruction. This method can also be used to study other dynamic geometry-related problems in particle and nuclear physics experiments, such as the expansion and contraction of the detector volumes owing to a temperature change, detector alignment at different running periods of the experiments, and geometry-related changes to the magnetic field.

	\subparagraph{Author contributions} All authors contributed to the study conception and design. Material preparation, data collection and analysis were performed by Shu Zhang, Jing-Shu Li, Yang-Jie Su, and Zi-Yuan Li. The first draft of the manuscript was written by Shu Zhang, Yu-Mei Zhang, and Zheng-Yun You, and all authors commented on previous versions of the manuscript. All authors read and approved the final manuscript.


\begin{thebibliography} {99}
		\bibitem{abe2013evidence} K. Abe, et al., Evidence for the Appearance of Atmospheric Tau Neutrinos in Super-Kamiokande. Phys. Rev. Lett. 110, 181802 (2013). \href{https://doi.org/10.1103/PhysRevLett.110.181802}{doi: 10.1103/PhysRevLett.110.181802}
		
		\bibitem{ahmad2002direct} Q. Ahmad, et al., Direct Evidence for Neutrino Flavor Transformation from Neutral-Current Interactions in the Sudbury Neutrino Observatory. Phys. Rev. Lett. 89, 011301 (2002). \href{https://doi.org/10.1103/PhysRevLett.89.011301}{doi: 10.1103/PhysRevLett.89.011301}
		
		\bibitem{adam2015juno} T. Adam, et al., JUNO Conceptual Design Report (2015). \href{https://arxiv.org/abs/1508.07166}{arXiv:1508.07166}
		
		\bibitem{Abe2018uyc} K. Abe, et al., Hyper-Kamiokande Design Report (2018). \href{https://arxiv.org/abs/1805.04163}{arXiv:1805.04163}
		
		\bibitem{acciarri2016long}	R. Acciarri, et al., Long-Baseline Neutrino Facility (LBNF) and Deep Underground Neutrino Experiment (DUNE) Conceptual Design Report, Volume 4 The DUNE Detectors at LBNF (2016). \href{https://arxiv.org/abs/1601.02984}{arXiv:1601.02984}
		
		\bibitem{cao2014liquid}X. Cao, X. Chen, Y. Chen, et al., PandaX: a liquid xenon dark matter experiment at CJPL. Sci. China Phys. Mech. Astron. 57, 1476-1494 (2014). \href{https://doi.org/10.1007/s11433-014-5521-2}{doi: 10.1007/s11433-014-5521-2}
		
		\bibitem{zhang2019dark} H. Zhang, et al., Dark matter direct search sensitivity of the PandaX-4T experiment. Sci. China Phys. Mech. Astron. 62, 31011 (2019). \href{https://doi.org/10.1007/s11433-018-9259-0}{doi: 10.1007/s11433-018-9259-0}
		
		\bibitem{akerib2017results} D.S. Akerib, et al., Results from a Search for Dark Matter in the Complete LUX Exposure. Phys. Rev. Lett. 118, 021303 (2017). \href{https://doi.org/10.1103/PhysRevLett.118.021303}{doi: 10.1103/PhysRevLett.118.021303}
		
		\bibitem{juyal2020proportional}P. Juyal, K.L. Giboni, X.D. Ji, et al., On proportional scintillation in very large liquid xenon detectors. Nucl. Sci. Tech. 31, 93 (2020). \href{https://doi.org/10.1007/s41365-020-00797-4}{doi: 10.1007/s41365-020-00797-4}
		
		\bibitem{huang2020simulation}M.Y. Huang, H. Pei, X.M. Sun, et al., Simulation study of energy resolution with changing pixel size for radon monitor based on \textit{Topmetal-${II}^-$}TPC. Nucl. Sci. Tech. 30, 16 (2019). \href{https://doi.org/10.1007/s41365-018-0532-8}{doi: 10.1007/s41365-018-0532-8}
		
		\bibitem{yan2020study}Y.L. Yan, W.X. Zhong, S.T. Lin, et al., Study on cosmogenic radioactive production in germanium as a background for future rare event search experiments. Nucl. Sci. Tech. 31, 55 (2020). \href{https://doi.org/10.1007/s41365-020-00762-1}{doi: 10.1007/s41365-020-00762-1}
		
		\bibitem{giboni2020ln2}K.L. Giboni, P. Juyal, E. Aprile, et al., A LN2-based cooling system for a next-generation liquid xenon dark matter detector. Nucl. Sci. Tech. 31, 76 (2020). \href{https://doi.org/10.1007/s41365-020-00786-7}{doi: 10.1007/s41365-020-00786-7}
		
		\bibitem{chytracek2006geometry} R. Chytracek, et al., Geometry Description Markup Language for Physics Simulation and Analysis Applications. IEEE Trans. Nucl. Sci. 53, 881062 (2006). \href{https://doi.org/10.1109/TNS.2006.881062}{doi: 10.1109/TNS.2006.881062}
		
		\bibitem{agostinelli2003geant4}	S. Agostinelli, et al., GEANT4$-$a simulation toolkit. Nucl. Instrum. Meth. A. 506, 250-303 (2003). \href{https://doi.org/10.1016/S0168-9002(03)01368-8}{doi: 10.1016/S0168-9002(03)01368-8}
		
		\bibitem{allison2006geant4} J. Allison, et al., Geant4 developments and applications. IEEE Trans. Nucl. Sci. 53, 869826 (2006). \href{https://doi.org/10.1109/TNS.2006.869826}{doi: 10.1109/TNS.2006.869826}
		
		\bibitem{multiphysics1998introduction} COMSOL Multiphysics, Introduction to COMSOL multiphysics. COMSOL Multiphysics. 2016. \href{https://www.comsol.com}{https://www.comsol.com}
		
		\bibitem{jasak2007openfoam} H. Jasak, A. Jemcov, Z. Tukovic, Openfoam: A C++ library for complex physics simulations. International Workshop on Coupled Methods in Numerical Dynamics, Dubrovnik, Croatia, 1–20 (2007). \href{http://cmnd2007.fsb.hr}{http://cmnd2007.fsb.hr}
		
		\bibitem{bray2000extensible} T. Bray, J. Paoli, C. Sperberg-McQueen, Extensible Markup Language (XML) 1.0. \href{http://www.w3.org/XML/1998/06/xmlspec-report-19980910.htm}{http://www.w3.org/XML/1998/06/xmlspec-report-19980910.htm}
		
		\bibitem{wang2020cad}X. Wang, J.L. Li, Z. Wu, et al., CMGC: a CAD to Monte Carlo geometry conversion code. Nucl. Sci. Tech. 31, 82 (2020). \href{https://doi.org/10.1007/s41365-020-00793-8}{doi: 10.1007/s41365-020-00793-8}
		
		\bibitem{you2008gdml} Z. You, et al., A method for detector description exchange among ROOT GEANT4 and GEANT3, Chin. Phys. C. 32, 572 (2008). \href{https://doi.org/10.1088/1674-1137/32/7/012}{doi: 10.1088/1674-1137/32/7/012}
		
		\bibitem{li2018gdml}	K. Li, et al., GDML based geometry management system for offline software in JUNO. Nucl. Instrum. Meth. A. 908, 43-48 (2018). \href{https://doi.org/10.1016/j.nima.2018.08.008}{doi: 10.1016/j.nima.2018.08.008}
		
		\bibitem{liang2009uniform}	Y. Liang, et al., A uniform geometry description for simulation, reconstruction and visualization in the BESIII experiment. Nucl. Instrum. Meth. A. 603, 325-327 (2009). \href{https://doi.org/10.1016/j.nima.2009.02.036}{doi: 10.1016/j.nima.2009.02.036}
		
		\bibitem{you2018root} Z. You, et al., A ROOT based event display software for JUNO. JINST, 13, T02002 (2018). \href{https://doi.org/10.1088/1748-0221/13/02/T02002}{doi: 10.1088/1748-0221/13/02/T02002}
		
		\bibitem{zhu2019method} J. Zhu, et al., A method of detector and event visualization with Unity in JUNO. JINST. 14, T01007 (2019). \href{https://doi.org/10.1088/1748-0221/14/01/T01007}{doi: 10.1088/1748-0221/14/01/T01007}
		
		\bibitem{dong2020study}H. Dong, D.Q. Fang, C. Li, Study on the performance of a large-size CsI detector for high energy $\gamma$-rays. Nucl. Sci. Tech. 29, 7 (2018). \href{https://doi.org/10.1007/s41365-017-0345-1}{doi: 10.1007/s41365-017-0345-1}
		
		\bibitem{an2013improved}	F.P. An, et al., Improved measurement of electron antineutrino disappearance at Daya Bay. Chin. Phys. C. 37, 011001 (2013). \href{https://doi.org/10.1088/1674-1137/37/1/011001}{doi: 10.1088/1674-1137/37/1/011001}
		
		\bibitem{abe2010production}	S. Abe, et al., Production of radioactive isotopes through cosmic muon spallation in KamLAND. Phys. Rev. C. 81, 025807 (2010). \href{https://doi.org/10.1103/PhysRevC.81.025807}{doi: 10.1103/PhysRevC.81.025807}
		
		\bibitem{amaudruz2019design} P.A. Amaudruz, et al., Design and construction of the DEAP-3600 dark matter detector. Astropart. Phys. 108, 1-23 (2019). \href{https://doi.org/10.1016/j.astropartphys.2018.09.006}{doi: 10.1016/j.astropartphys.2018.09.006}
		
		\bibitem{aprile2019light} E. Aprile, et al., Light dark matter search with ionization signals in XENON1T. Phys. Rev. Lett. 123, 251801 (2019). \href{https://doi.org/10.1103/PhysRevLett.123.251801}{doi: 10.1103/PhysRevLett.123.251801}
		
		\bibitem{agnes2018low} P. Agnes, et al., Low-Mass Dark Matter Search with the DarkSide-50 Experiment. Phys. Rev. Lett. 121, 081307 (2018). \href{https://doi.org/10.1103/PhysRevLett.121.081307}{doi: 10.1103/PhysRevLett.121.081307}
		
		\bibitem{schiebener1990refractive} P. Schiebener, et al., Refractive index of water and steam as function of wavelength, temperature and density. J. Phys. Chem. Ref. Data 19, 677 (1990). \href{https://doi.org/10.1063/1.555859}{doi: 10.1063/1.555859}
		
		\bibitem{brun1997root} R. Brun, F. Rademakers, ROOT - An object oriented data analysis framework. Nucl. Instrum. Meth. A. 389, S0168 (1997). \href{https://doi.org/10.1016/S0168-9002(97)00048-X}{doi: 10.1016/S0168-9002(97)00048-X}
		
		
	\end{thebibliography}
	
\end{document}